\documentclass[aps,prd,twocolumn,amsmath,amssymb,amsfonts,nofootinbib,superscriptaddress,altaffilletter]{revtex4-1}
\usepackage{array}
\newcolumntype{P}[1]{>{\centering\arraybackslash}p{#1}}

\usepackage{graphicx}
\usepackage{amssymb}
\usepackage{amsmath}
\usepackage{longtable}
\usepackage{color}
\usepackage{etoolbox}
\usepackage{supertabular}
\usepackage{dcolumn}
\usepackage{ wasysym }

\newcommand{\maxfdot}{2.6 $\times 10^{-9}$ Hz/s}

\def\Tobs{T_{\textrm{\mbox{\tiny{obs}}}}}
\def\Tcoh{T_{\textrm{\mbox{\tiny{coh}}}}}
\def\Tref{T_{\textrm{\mbox{\tiny{ref}}}}}

\def\EatH{Einstein@Home}
\def\EatHs{Einstein@Home }

\newcommand{\ee}[1]{\!\times\!10^{#1}}

\def\sci#1#2{#1\times10^{#2}}

\newcommand{\avgSeg}[1]{\overline{#1}}			

\newcommand{\Gauss}{\mathrm{\MakeUppercase{G}}}
\newcommand{\Signal}{{\mathrm{\MakeUppercase{S}}}}
\newcommand{\Line}{{\mathrm{\MakeUppercase{L}}}}
\newcommand{\Transient}{{\mathrm{t\MakeUppercase{L}}}}
\newcommand{\Noise}{{\Gauss\Line}}
\newcommand{\NoisetL}{{\Gauss\Line\Transient}}

\providecommand{\sc}[1]{\widehat{#1}}
\renewcommand{\sc}[1]{\widehat{#1}}

\newcommand{\Bayes}{\hat{B}}

\newcommand{\BSNtsc}{{\hat\beta}_{{\Signal/\NoisetL}}}	
\newcommand{\notlogBSNtsc}{\Bayes_{{\Signal/\NoisetL}}}
\newcommand{\notlogBSNsc}{\Bayes_{{\Signal/\Noise}}}	
\newcommand{\BSNsc}{{\hat\beta}_{{\Signal/\Noise}}}	

\newcommand{\F}{\mathcal{F}}		

\newcommand{\scF}{\sc{\F}}

\newcommand{\scFtho}{\scF_*^{(0)}}

\newcommand{\avF}{\avgSeg{\F}}

\newcommand{\Nseg}{{N_{\mathrm{seg}}}}

\newcommand{\paramfdotlo}{\ensuremath{-\sci{2.65}{-9}~\mathrm{Hz/s}}} 
\newcommand{\paramfdothi}{\ensuremath{\sci{2.64}{-10}~\mathrm{Hz/s}}} 
\newcommand{\paramWUcputimeHours}{8} 
\newcommand{\paramWUavgskypointsInt}{118} 
\newcommand{\paramWUtotaltemplates}{\ensuremath{\sci{1.5}{11}}} 
\newcommand{\paramtotalWUsmillions}{1.9} 
\newcommand{\paramtotaltemplates}{\ensuremath{\sci{3}{17}}} 

\newcommand{\sensDepth}{48.7~}
\newcommand{\lowestUL}{1.8 \times 10^{-25}}

\newcommand{\highestUL}{3.9 \times 10^{-24}}

\newtoggle{fullauthorlist}
\toggletrue{fullauthorlist}

\newtoggle{endauthorlist}
\toggletrue{endauthorlist}

\begin{document}

\title{ 
First low-frequency Einstein@Home all-sky search for continuous gravitational waves in Advanced LIGO data
}


\iftoggle{endauthorlist}{
  %
  %
  \let\mymaketitle\maketitle
  \let\myauthor\author
  \let\myaffiliation\affiliation
  \author{The LIGO Scientific Collaboration}
  \author{The Virgo Collaboration {\it et al.}} 
  
}{
  %
  %
  \iftoggle{fullauthorlist}{
    \author{%
B.~P.~Abbott,$^{1}$  
R.~Abbott,$^{1}$  
T.~D.~Abbott,$^{2}$  
F.~Acernese,$^{3,4}$ 
K.~Ackley,$^{5}$  
C.~Adams,$^{6}$  
T.~Adams,$^{7}$ 
P.~Addesso,$^{8}$  
R.~X.~Adhikari,$^{1}$  
V.~B.~Adya,$^{9}$
% ...
\\
\medskip
(LIGO Scientific Collaboration and Virgo Collaboration) 
\\
\medskip
and D.~P.~Anderson$^{147}$
\\
\medskip
{{}$^{*}$Deceased, March 2016. }%
{{}$^{\sharp}$Deceased, March 2017. }%
{${}^{\dag}$Deceased, February 2017. }%
{${}^{\ddag}$Deceased, December 2016. }%
}\noaffiliation
\affiliation {LIGO, California Institute of Technology, Pasadena, CA 91125, USA }
\affiliation {Louisiana State University, Baton Rouge, LA 70803, USA }
\affiliation {Universit\`a di Salerno, Fisciano, I-84084 Salerno, Italy }
\affiliation {INFN, Sezione di Napoli, Complesso Universitario di Monte S.Angelo, I-80126 Napoli, Italy }
\affiliation {University of Florida, Gainesville, FL 32611, USA }
\affiliation {LIGO Livingston Observatory, Livingston, LA 70754, USA }
\affiliation {Laboratoire d'Annecy-le-Vieux de Physique des Particules (LAPP), Universit\'e Savoie Mont Blanc, CNRS/IN2P3, F-74941 Annecy, France }
\affiliation {University of Sannio at Benevento, I-82100 Benevento, Italy and INFN, Sezione di Napoli, I-80100 Napoli, Italy }
\affiliation {Albert-Einstein-Institut, Max-Planck-Institut f\"ur Gravi\-ta\-tions\-physik, D-30167 Hannover, Germany }
\affiliation {The University of Mississippi, University, MS 38677, USA }
\affiliation {NCSA, University of Illinois at Urbana-Champaign, Urbana, IL 61801, USA }
\affiliation {University of Cambridge, Cambridge CB2 1TN, United Kingdom }
\affiliation {Nikhef, Science Park, 1098 XG Amsterdam, The Netherlands }
\affiliation {LIGO, Massachusetts Institute of Technology, Cambridge, MA 02139, USA }
\affiliation {Instituto Nacional de Pesquisas Espaciais, 12227-010 S\~{a}o Jos\'{e} dos Campos, S\~{a}o Paulo, Brazil }
\affiliation {Gran Sasso Science Institute (GSSI), I-67100 L'Aquila, Italy }
\affiliation {INFN, Sezione di Roma Tor Vergata, I-00133 Roma, Italy }
\affiliation {Inter-University Centre for Astronomy and Astrophysics, Pune 411007, India }
\affiliation {University of Wisconsin-Milwaukee, Milwaukee, WI 53201, USA }
\affiliation {Leibniz Universit\"at Hannover, D-30167 Hannover, Germany }
\affiliation {Universit\`a di Pisa, I-56127 Pisa, Italy }
\affiliation {INFN, Sezione di Pisa, I-56127 Pisa, Italy }
\affiliation {OzGrav, Australian National University, Canberra, Australian Capital Territory 0200, Australia }
\affiliation {Laboratoire des Mat\'eriaux Avanc\'es (LMA), CNRS/IN2P3, F-69622 Villeurbanne, France }
\affiliation {LAL, Univ. Paris-Sud, CNRS/IN2P3, Universit\'e Paris-Saclay, F-91898 Orsay, France }
\affiliation {California State University Fullerton, Fullerton, CA 92831, USA }
\affiliation {European Gravitational Observatory (EGO), I-56021 Cascina, Pisa, Italy }
\affiliation {Universit\`a di Roma Tor Vergata, I-00133 Roma, Italy }
\affiliation {Universit\"at Hamburg, D-22761 Hamburg, Germany }
\affiliation {INFN, Sezione di Roma, I-00185 Roma, Italy }
\affiliation {Embry-Riddle Aeronautical University, Prescott, AZ 86301, USA }
\affiliation {Albert-Einstein-Institut, Max-Planck-Institut f\"ur Gravitations\-physik, D-14476 Potsdam-Golm, Germany }
\affiliation {APC, AstroParticule et Cosmologie, Universit\'e Paris Diderot, CNRS/IN2P3, CEA/Irfu, Observatoire de Paris, Sorbonne Paris Cit\'e, F-75205 Paris Cedex 13, France }
\affiliation {Korea Institute of Science and Technology Information, Daejeon 34141, Korea }
\affiliation {West Virginia University, Morgantown, WV 26506, USA }
\affiliation {Center for Gravitational Waves and Cosmology, West Virginia University, Morgantown, WV 26505, USA }
\affiliation {Universit\`a di Perugia, I-06123 Perugia, Italy }
\affiliation {INFN, Sezione di Perugia, I-06123 Perugia, Italy }
\affiliation {Syracuse University, Syracuse, NY 13244, USA }
\affiliation {University of Minnesota, Minneapolis, MN 55455, USA }
\affiliation {SUPA, University of Glasgow, Glasgow G12 8QQ, United Kingdom }
\affiliation {LIGO Hanford Observatory, Richland, WA 99352, USA }
\affiliation {Wigner RCP, RMKI, H-1121 Budapest, Konkoly Thege Mikl\'os \'ut 29-33, Hungary }
\affiliation {Columbia University, New York, NY 10027, USA }
\affiliation {Stanford University, Stanford, CA 94305, USA }
\affiliation {Universit\`a di Camerino, Dipartimento di Fisica, I-62032 Camerino, Italy }
\affiliation {Universit\`a di Padova, Dipartimento di Fisica e Astronomia, I-35131 Padova, Italy }
\affiliation {INFN, Sezione di Padova, I-35131 Padova, Italy }
\affiliation {MTA E\"otv\"os University, ``Lendulet'' Astrophysics Research Group, Budapest 1117, Hungary }
\affiliation {Nicolaus Copernicus Astronomical Center, Polish Academy of Sciences, 00-716, Warsaw, Poland }
\affiliation {University of Birmingham, Birmingham B15 2TT, United Kingdom }
\affiliation {Universit\`a degli Studi di Genova, I-16146 Genova, Italy }
\affiliation {INFN, Sezione di Genova, I-16146 Genova, Italy }
\affiliation {RRCAT, Indore MP 452013, India }
\affiliation {Faculty of Physics, Lomonosov Moscow State University, Moscow 119991, Russia }
\affiliation {SUPA, University of the West of Scotland, Paisley PA1 2BE, United Kingdom }
\affiliation {Caltech CaRT, Pasadena, CA 91125, USA }
\affiliation {OzGrav, University of Western Australia, Crawley, Western Australia 6009, Australia }
\affiliation {Department of Astrophysics/IMAPP, Radboud University Nijmegen, P.O. Box 9010, 6500 GL Nijmegen, The Netherlands }
\affiliation {Artemis, Universit\'e C\^ote d'Azur, Observatoire C\^ote d'Azur, CNRS, CS 34229, F-06304 Nice Cedex 4, France }
\affiliation {Institut de Physique de Rennes, CNRS, Universit\'e de Rennes 1, F-35042 Rennes, France }
\affiliation {Washington State University, Pullman, WA 99164, USA }
\affiliation {Universit\`a degli Studi di Urbino 'Carlo Bo', I-61029 Urbino, Italy }
\affiliation {INFN, Sezione di Firenze, I-50019 Sesto Fiorentino, Firenze, Italy }
\affiliation {University of Oregon, Eugene, OR 97403, USA }
\affiliation {Laboratoire Kastler Brossel, UPMC-Sorbonne Universit\'es, CNRS, ENS-PSL Research University, Coll\`ege de France, F-75005 Paris, France }
\affiliation {Carleton College, Northfield, MN 55057, USA }
\affiliation {Astronomical Observatory Warsaw University, 00-478 Warsaw, Poland }
\affiliation {VU University Amsterdam, 1081 HV Amsterdam, The Netherlands }
\affiliation {University of Maryland, College Park, MD 20742, USA }
\affiliation {Center for Relativistic Astrophysics and School of Physics, Georgia Institute of Technology, Atlanta, GA 30332, USA }
\affiliation {Universit\'e Claude Bernard Lyon 1, F-69622 Villeurbanne, France }
\affiliation {Universit\`a di Napoli 'Federico II', Complesso Universitario di Monte S.Angelo, I-80126 Napoli, Italy }
\affiliation {NASA Goddard Space Flight Center, Greenbelt, MD 20771, USA }
\affiliation {RESCEU, University of Tokyo, Tokyo, 113-0033, Japan. }
\affiliation {OzGrav, University of Adelaide, Adelaide, South Australia 5005, Australia }
\affiliation {Tsinghua University, Beijing 100084, China }
\affiliation {Texas Tech University, Lubbock, TX 79409, USA }
\affiliation {Kenyon College, Gambier, OH 43022, USA }
\affiliation {The Pennsylvania State University, University Park, PA 16802, USA }
\affiliation {National Tsing Hua University, Hsinchu City, 30013 Taiwan, Republic of China }
\affiliation {Charles Sturt University, Wagga Wagga, New South Wales 2678, Australia }
\affiliation {University of Chicago, Chicago, IL 60637, USA }
\affiliation {Pusan National University, Busan 46241, Korea }
\affiliation {The Chinese University of Hong Kong, Shatin, NT, Hong Kong }
\affiliation {INAF, Osservatorio Astronomico di Padova, Vicolo dell'Osservatorio 5, I-35122 Padova, Italy }
\affiliation {INFN, Trento Institute for Fundamental Physics and Applications, I-38123 Povo, Trento, Italy }
\affiliation {Universit\`a di Roma 'La Sapienza', I-00185 Roma, Italy }
\affiliation {Universit\'e Libre de Bruxelles, Brussels 1050, Belgium }
\affiliation {Sonoma State University, Rohnert Park, CA 94928, USA }
\affiliation {Montana State University, Bozeman, MT 59717, USA }
\affiliation {Center for Interdisciplinary Exploration \& Research in Astrophysics (CIERA), Northwestern University, Evanston, IL 60208, USA }
\affiliation {Cardiff University, Cardiff CF24 3AA, United Kingdom }
\affiliation {Universitat de les Illes Balears, IAC3---IEEC, E-07122 Palma de Mallorca, Spain }
\affiliation {The University of Texas Rio Grande Valley, Brownsville, TX 78520, USA }
\affiliation {Bellevue College, Bellevue, WA 98007, USA }
\affiliation {Institute for Plasma Research, Bhat, Gandhinagar 382428, India }
\affiliation {The University of Sheffield, Sheffield S10 2TN, United Kingdom }
\affiliation {University of Southern California Information Sciences Institute, Marina Del Rey, CA 90292, USA }
\affiliation {California State University, Los Angeles, 5151 State University Dr, Los Angeles, CA 90032, USA }
\affiliation {Universit\`a di Trento, Dipartimento di Fisica, I-38123 Povo, Trento, Italy }
\affiliation {Montclair State University, Montclair, NJ 07043, USA }
\affiliation {National Astronomical Observatory of Japan, 2-21-1 Osawa, Mitaka, Tokyo 181-8588, Japan }
\affiliation {Canadian Institute for Theoretical Astrophysics, University of Toronto, Toronto, Ontario M5S 3H8, Canada }
\affiliation {Whitman College, 345 Boyer Avenue, Walla Walla, WA 99362 USA }
\affiliation {School of Mathematics, University of Edinburgh, Edinburgh EH9 3FD, United Kingdom }
\affiliation {University and Institute of Advanced Research, Gandhinagar Gujarat 382007, India }
\affiliation {IISER-TVM, CET Campus, Trivandrum Kerala 695016, India }
\affiliation {University of Szeged, D\'om t\'er 9, Szeged 6720, Hungary }
\affiliation {International Centre for Theoretical Sciences, Tata Institute of Fundamental Research, Bengaluru 560089, India }
\affiliation {Tata Institute of Fundamental Research, Mumbai 400005, India }
\affiliation {INAF, Osservatorio Astronomico di Capodimonte, I-80131, Napoli, Italy }
\affiliation {University of Michigan, Ann Arbor, MI 48109, USA }
\affiliation {American University, Washington, D.C. 20016, USA }
\affiliation {Rochester Institute of Technology, Rochester, NY 14623, USA }
\affiliation {University of Bia{\l }ystok, 15-424 Bia{\l }ystok, Poland }
\affiliation {SUPA, University of Strathclyde, Glasgow G1 1XQ, United Kingdom }
\affiliation {University of Southampton, Southampton SO17 1BJ, United Kingdom }
\affiliation {University of Washington Bothell, 18115 Campus Way NE, Bothell, WA 98011, USA }
\affiliation {Institute of Applied Physics, Nizhny Novgorod, 603950, Russia }
\affiliation {Seoul National University, Seoul 08826, Korea }
\affiliation {Inje University Gimhae, South Gyeongsang 50834, Korea }
\affiliation {National Institute for Mathematical Sciences, Daejeon 34047, Korea }
\affiliation {NCBJ, 05-400 \'Swierk-Otwock, Poland }
\affiliation {Institute of Mathematics, Polish Academy of Sciences, 00656 Warsaw, Poland }
\affiliation {OzGrav, School of Physics \& Astronomy, Monash University, Clayton 3800, Victoria, Australia }
\affiliation {Hanyang University, Seoul 04763, Korea }
\affiliation {NASA Marshall Space Flight Center, Huntsville, AL 35811, USA }
\affiliation {ESPCI, CNRS, F-75005 Paris, France }
\affiliation {Southern University and A\&M College, Baton Rouge, LA 70813, USA }
\affiliation {OzGrav, University of Melbourne, Parkville, Victoria 3010, Australia }
\affiliation {College of William and Mary, Williamsburg, VA 23187, USA }
\affiliation {Indian Institute of Technology Madras, Chennai 600036, India }
\affiliation {IISER-Kolkata, Mohanpur, West Bengal 741252, India }
\affiliation {Scuola Normale Superiore, Piazza dei Cavalieri 7, I-56126 Pisa, Italy }
\affiliation {Universit\'e de Lyon, F-69361 Lyon, France }
\affiliation {Hobart and William Smith Colleges, Geneva, NY 14456, USA }
\affiliation {Janusz Gil Institute of Astronomy, University of Zielona G\'ora, 65-265 Zielona G\'ora, Poland }
\affiliation {University of Washington, Seattle, WA 98195, USA }
\affiliation {King's College London, University of London, London WC2R 2LS, United Kingdom }
\affiliation {Indian Institute of Technology, Gandhinagar Ahmedabad Gujarat 382424, India }
\affiliation {International Institute of Physics, Universidade Federal do Rio Grande do Norte, Natal RN 59078-970, Brazil }
\affiliation {Andrews University, Berrien Springs, MI 49104, USA }
\affiliation {Universit\`a di Siena, I-53100 Siena, Italy }
\affiliation {Trinity University, San Antonio, TX 78212, USA }
\affiliation {Abilene Christian University, Abilene, TX 79699, USA }
\affiliation {University of California at Berkeley, Berkeley, CA 94720 USA }

  }{
    \author{The LIGO Scientific Collaboration}
    \affiliation{LSC}
    \author{The Virgo Collaboration}
    \affiliation{Virgo}
  }
}

\begin{abstract}
\iftoggle{endauthorlist}{
}{
  \newpage
}
We report results of a deep all-sky search for periodic gravitational waves from isolated neutron stars in data from the first Advanced LIGO observing run. This search investigates the low frequency range of Advanced LIGO data, between 20 and 100 Hz, much of which was not explored in initial LIGO. The search was made possible by the computing power provided by the volunteers of the \EatHs project. We find no significant signal candidate and set the most stringent upper limits to date on the amplitude of gravitational wave signals from the target population, corresponding to a sensitivity depth of $48.7$ [1/$\sqrt{{\textrm{Hz}}}$]. 
At the frequency of best strain sensitivity, near $100$ Hz, we set 90\%\ confidence upper limits of $\lowestUL$. At the low end of our frequency range, $20$ Hz, we achieve upper limits of $\highestUL$. At $55$ Hz we can exclude sources with ellipticities greater than $10^{-5}$ within 100 pc of Earth with fiducial value of the principal moment of inertia of $10^{38} \textrm{kg m}^2$. 
\\
\end{abstract}

\maketitle

\author{The LIGO Scientific Collaboration and the Virgo Collaboration}
\thanks{Full author list given at the end of the article.}

\maketitle

\section{Introduction}
\label{sec:introduction}

In this paper we report the results of a deep all-sky \EatHs \cite{EaHweb} search for continuous, nearly monochromatic gravitational waves (GWs) in data from the first Advanced LIGO observing run (O1). A number of all-sky searches have been carried out on initial LIGO data, \cite{S6EHFU,S6EH,S6Powerflux, Aasi:2015rar, S5GC1HF, FullS5EH,FullS5Semicoherent, S5EH, EarlyS5Paper,S5SkyHough,VSR1TDFstat,S4IncoherentPaper,S4EH,S2FstatPaper}, of which \cite{S4EH, S5EH, FullS5EH, S6EH, S6EHFU} also ran on \EatH. \EatH~is a distributed computing project which uses the idle time of computers volunteered by the general public to search for GWs.

The search presented here covers frequencies from 20~Hz through 100~Hz and frequency derivatives from \paramfdotlo~through {\paramfdothi}. A large portion of this frequency range was not explored in initial LIGO due to lack of sensitivity. By focusing the available computing power on a subset of the detector frequency range, this search achieves higher sensitivity at these low frequencies than would be possible in a search over the full range of LIGO frequencies.  In this low-frequency range we establish the most constraining gravitational wave amplitude upper limits to date for the target signal population.

\section{LIGO interferometers and the data used}
\label{sec:S6intro} 

The LIGO gravitational wave network consists of two observatories, one in Hanford (WA) and the other in Livingston (LA) separated by a 3000-km baseline \cite{LIGO_detector}. The first observing run (O1) \cite{LIGO_O1} of this network after the upgrade towards the Advanced LIGO configuration \cite{TheLIGOScientific:2014jea} took place between September 2015 and January 2016. The Advanced LIGO detectors are significantly more sensitive than the initial LIGO detectors. This increase in sensitivity is especially significant in the low-frequency range of 20~Hz through 100~Hz covered by this search: at 100~Hz the O1 Advanced LIGO detectors are about a factor 5 more sensitive than the Initial LIGO detectors during their last run (S6 \cite{LIGO:2012aa}), and this factor becomes $\approx$ 20 at 50~Hz. For this reason all-sky searches did not include frequencies below 50~Hz on initial LIGO data.

Since interferometers sporadically fall out of operation (``lose lock'') due to environmental or instrumental disturbances or for scheduled maintenance periods, the data set is not contiguous and each detector has a duty factor of about 50\%. To remove the effects of instrumental and environmental spectral disturbances from the analysis, the data in frequency bins known to contain such disturbances have been substituted with Gaussian noise with the same average power as that in the neighbouring and undisturbed bands. This is the same procedure as used in \cite{S6EH}. These bands are identified in the Appendix.

\section{The Search}
\label{sec:search}

The search described in this paper targets nearly monochromatic gravitational wave signals as described for example by Eqs. 1-4 of \cite{S5EH}. Various emission mechanisms could generate such a signal, as reviewed in Section IIA of \cite{S2FstatPaper}. In interpreting our results we will consider a spinning compact object with a fixed, non-axisymmetric $\ell=m=2$ mass quadrupole, described by an equatorial ellipticity $\epsilon$.

We perform a stack-slide type of search using the GCT (Global correlation transform) method \cite{PletschAllen,Pletsch:2008,Pletsch:2010}. In a stack-slide search the data is partitioned in segments, and each segment is searched with a matched-filter method \cite{cutler}. The results from these coherent searches are combined by summing the detection statistic values from the different segments, one per segment ($\F_i$), and this determines the value of the core detection statistic: 
\begin{equation}
\label{eq:avF}
\avF:={1\over\Nseg} \sum_{i=1}^{\Nseg} \F_i.
\end{equation}
The ``stacking'' part of the procedure is the summing and the ``sliding'' (in parameter space) refers to the fact that the $\F_i$ that are summed do not all come from the same template. 

Summing the detection statistic values is not the only way to combine the results from the coherent searches, see for instance \cite{S6Powerflux,HoughMethods2004,FH_2}. Independently of the way that this is done, this type of search is usually referred to as a ``semi-coherent search''. Important variables for this type of search are: the coherent time baseline of the segments $\Tcoh$, the number of segments used $\Nseg$, the total time spanned by the data $\Tobs$, the grids in parameter space and the detection statistic used to rank the parameter space cells. For a stack-slide search in Gaussian noise, $\Nseg\times 2\avF$ follows a chi-squared distribution with $4\Nseg$ degrees of freedom, $\chi^2_{4\Nseg}$.
These parameters are summarised in Table \ref{tab:GridSpacings}. The grids in frequency and spindown are each described by a single parameter, the grid spacing, which is constant over the search range. The same frequency grid spacings are used for the coherent searches over the segments and for the incoherent summing. The spindown spacing for the incoherent summing, $\delta{\dot{f}}$, is finer than that used for the coherent searches, $\delta{\dot{f_c}}$, by a factor $\gamma$. The notation used here is consistent with that used in previous observational papers \cite{S6EHFU, S6EH}.

\begin{figure}[h!tbp]
   \includegraphics[width=\columnwidth]{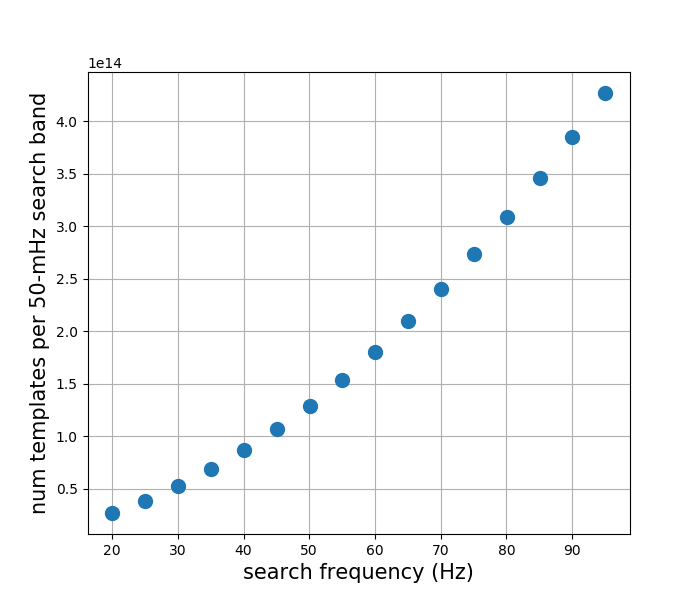}
\caption{Number of searched templates in 50\,mHz band as a function of frequency. The sky resolution increases with frequency causing the variation in the number of templates. 
$N_f \times N_{\dot{f}} \sim 1.3\times 10^{9}$, where $N_f$ and $N_{\dot{f}}$ are the number of $f$ and $\dot{f}$ templates searched in 50\,mHz bands. The total number of templates searched between 20 and 100 Hz is {\paramtotaltemplates}. 
}  
\label{fig:NumberOfTemplatesIn50mHz}
\end{figure}

The sky grid is approximately uniform on the celestial sphere projected on the ecliptic plane. The tiling is an hexagonal covering of the unit circle with hexagons' edge length $d$:
\begin{equation}
d(m_{\text{sky}})={1\over f}
{
{\sqrt{ m_{\text{sky}} } }
\over {\pi \tau_{E}}
} ,
\label{eq:skyGridSpacing}
\end{equation}
with $\tau_{E}\simeq0.021$ s being half of the light travel time across the Earth and $m_{\text{sky}}$ a constant which controls the resolution of the sky grid. The sky-grids are constant over 5\,Hz bands and the spacings are the ones associated through Eq.~\ref{eq:skyGridSpacing} to the highest frequency in each 5\,Hz. The resulting number of templates used to search 50 mHz bands as a function of frequency is shown in Fig. \ref{fig:NumberOfTemplatesIn50mHz}. 

\begin{table}[t]
\begin{tabular}{|c|c|}
\hline
\hline
Parameter & Value \\
\hline
\hline
 $\Tcoh$ & 210 hr\\
 \hline
 $\Tref$  & 1132729647.5 GPS s \\
  \hline
$\Nseg$ & 12 \\
  \hline
$\delta f$ & $8.3\times 10^{-7}$ Hz \\
 \hline
 $\delta {\dot{f_c}}$ & $ 1.3\times 10^{-11}$ Hz/s \\
  \hline
$\gamma$ & 100 \\
  \hline
$m_{\text{sky}}$ &$1\ee{-3}$\\
 \hline
\hline
\end{tabular}

\caption{Search parameters rounded to the first decimal figure. $\Tref$ is the reference time that defines the frequency and frequency derivative values.
}
\label{tab:GridSpacings}
\end{table}

This search leverages the computing power of the \EatH~project, which is built upon the BOINC (Berkeley Open Infrastructure for Network Computing) architecture~\cite{Boinc1,Boinc2,Boinc3}: a system that exploits the idle time on volunteer computers to solve scientific problems that require large amounts of computer power. The search is split into work-units (WUs) sized to keep the average \EatHs volunteer computer busy for about {\paramWUcputimeHours} CPU-hours. Each WU performs {\paramWUtotaltemplates}~ semi-coherent searches, one for each of the templates in $50$ mHz band, the entire spindown range and {\paramWUavgskypointsInt} points in the sky. Out of the semicoherent detection statistic values computed for the {\paramWUtotaltemplates}~ templates, it returns to the \EatHs server only the highest 10000 values. A total of {\paramtotalWUsmillions}~million WUs are necessary to cover the entire parameter space. 
The total number of templates searched is \paramtotaltemplates.

\subsection{The ranking statistic}
Two detection statistics are used in the search: $\BSNtsc$ and $2\avF$. $\BSNtsc$ is the ranking statistic which defines the top-candidate-list; it is a line- and transient-robust statistic that tests the signal hypothesis against a noise model which, in addition to Gaussian noise, also includes single-detector continuous or transient spectral lines. Since the distribution of $\BSNtsc$ is not known in closed form even in Gaussian noise, when assessing the significance of a candidate against Gaussian noise, we use the average $2\F$ statistic over the segments, $2\avF$ \cite{cutler}, see Eq. \ref{eq:avF}. This is in essence, at every template point, the log-likelihood of having a signal with the shape given by the template versus having Gaussian noise.

Built from the multi- and single-detector $\scF$-statistics, $\BSNtsc$ is the log$_{10}$ of $\notlogBSNtsc$, the full definition of which is given by Eq.~(23) of~\cite{Keitel:2016}. This statistic depends on a few tuning parameters that we describe in the remainder of the paragraph for the reader interested in the technical details: A transition-scale parameter $\scFtho$ is used to tune the behaviour of
the $\BSNtsc$ statistic to match the performance of the standard average $2\avF$ statistic in Gaussian noise
while still statistically outperforming it in the presence of
continuous or transient single-detector spectral disturbances.
Based on injection studies of fake signals in Gaussian-noise data,
we set an average $2\avF$ transition scale of $\scFtho = 65.826$.
According to Eq.~67 of~\cite{Keitel:2013},
with $\Nseg = 12$ this $2\avF$ value corresponds to a Gaussian false-alarm probability of $10^{-9}$.
Furthermore, we assume equal-odds priors between the various noise hypotheses (``L'' for line, ``G" for Gaussian, ``tL" for transient-line).

\subsection{Identification of undisturbed bands}
\label{sec:visualInspection}

Even after the removal of disturbed data caused by spectral artefacts of known origin, the statistical properties of the results are not uniform across the search band. In what follows we concentrate on the subset of the signal-frequency bands having reasonably uniform statistical properties, or containing features that are not immediately identifiable as detector artefacts. This comprises the large majority of the search parameter space.

Our classification of ``clean" vs. ``disturbed" bands has no pretence of being strictly rigorous, because strict rigour here is neither useful nor practical. The classification serves the practical purpose of discarding from the analysis regions in parameter space with evident disturbances and must not dismiss detectable real signals. The classification is carried out in two steps: an automated identification of undisturbed bands and a visual inspection of the remaining bands. 

An automatic procedure, described in Section IIF of \cite{S6EHCasA}, identifies as undisturbed the 50-mHz bands whose maximum density of outliers in the $f-\dot{f}$ plane and average $2\avF$ are well within the bulk distribution of the values for these quantities in the neighbouring frequency bands. This procedure identifies $1233$ of the $1600$ 50-mHz bands as undisturbed. The remaining $367$ bands are marked as potentially disturbed, and in need of visual inspection. 

A scientist performs the visual inspection by looking at various distributions of the $\BSNtsc$ statistic over the entire sky and spindown parameter space in the $367$ potentially disturbed 50-mHz bands. She ranks each band with an integer score 0,1,2 ranging from ``undisturbed" (0) to ``disturbed" (2) . A band is considered ``undisturbed" if the distribution of detection statistic values does not show a visible trend affecting a large portion of the $f-\dot{f}$ plane. A band is considered ``mildly disturbed'' if there are outliers in the band that are localised in a small region of the $f-\dot{f}$ plane. A band is considered ``disturbed'' if there are outliers that are not well localised in the $f-\dot{f}$ plane.

\begin{figure}[h!tbp]
  \includegraphics[width=3.5in,height=3in]{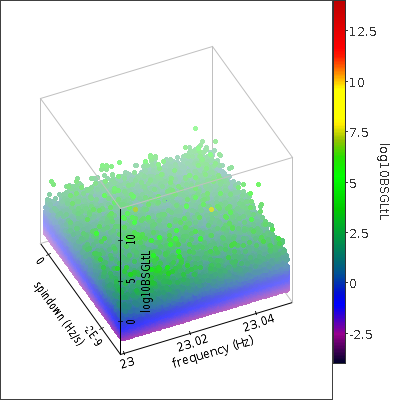}
  \includegraphics[width=3.5in,height=3in]{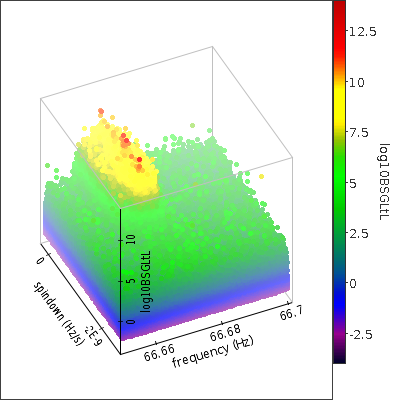}
   \includegraphics[width=3.5in,height=3in]{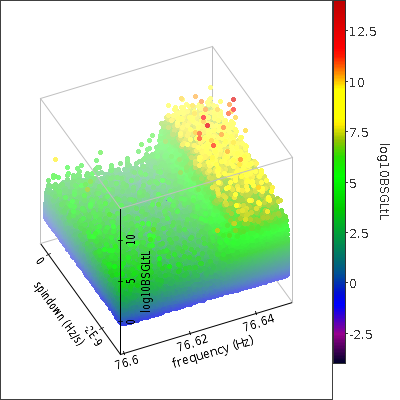}

\caption{On the vertical axis and color-coded is the $\BSNtsc$ in three 50-mHz bands. The top band was marked as ``undisturbed". The middle band is an example of a ``mildly disturbed band". The bottom band is an example of a ``disturbed band".}

\label{fig:VIGquietdisturbed}
\end{figure}

\begin{figure}[h!tbp]
   \includegraphics[width=3.5in,height=3in]{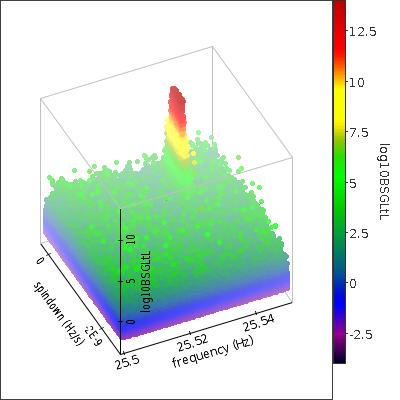}
\caption{This is an example of an ``undisturbed band" but containing a fake signal. On the z-axis and color-coded is the $\BSNtsc$.}
\label{fig:VIGsignal}
\end{figure}

Fig. \ref{fig:VIGquietdisturbed} shows the $\BSNtsc$ for each type of band. Fig. \ref{fig:VIGsignal} shows the $\BSNtsc$ for a band that harbours a fake signal injected in the data to verify the detection pipelines. In the latter case, the detection statistic is elevated in a small region around the signal parameters. 

Based on this visual inspection, 1\% of the bands between 20 and 100 Hz are marked as ``disturbed" and excluded from the current analysis. A further 6\% of the bands are marked as ``mildly disturbed''. These bands contain features that can not be classified as detector disturbances without further study, therefore these are included in the analysis.

\begin{figure}[h!tbp] 
\includegraphics[width=\columnwidth]{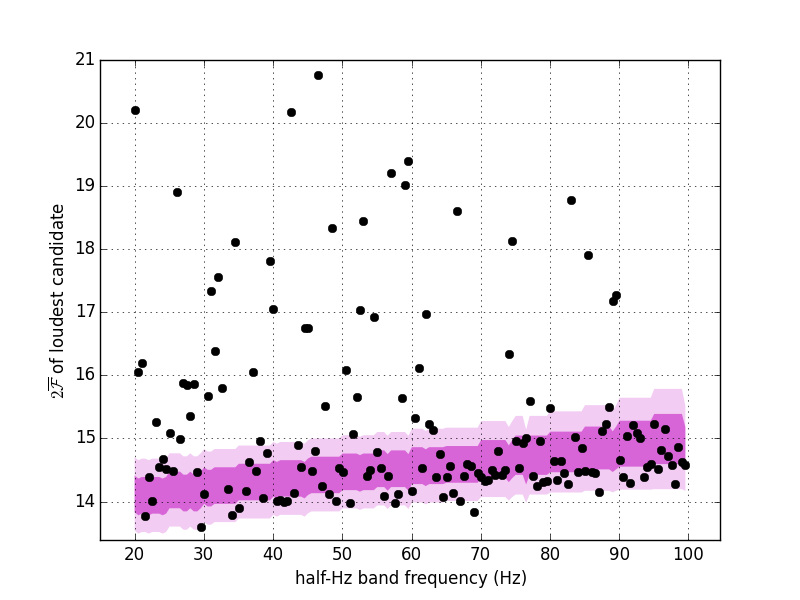}
        \caption{Highest $2\avF$ value (also referred to as the $2\avF$ of the loudest candidate) in every half-Hz band as a function of band frequency. Since the number of templates increases with frequency, so does the highest $2\avF$. The highest expected $2\avF \pm 1 \sigma (2 \sigma)$ over $N_{\text{trials}}$ independent trials is indicated by the darker (faded) band. Two half-mHz bands have $2\avF$ values greater than the axes boundaries. The half-Hz bands beginning at 33.05\,Hz and 35.55\,Hz have loudest $2\avF$ values of 159 and 500 respectively, due to features in the 33.3\,Hz and 35.75\,Hz 50-mHz bands which were marked ``mildly disturbed'' in the visual inspection.}
\label{fig:loudestHalfHzVersusFreq}
\end{figure}

Fig. \ref{fig:loudestHalfHzVersusFreq} shows the highest values of the detection statistic in half-Hz signal-frequency bands compared to the expectations. The set of candidates from which the highest detection statistic values are picked, does not include the 50-mHz signal-frequency bands that stem entirely from fake data, from the cleaning procedure, or that were marked as disturbed. Two 50-mHz bands that contained a hardware injection \cite{Biwer:2016oyg} were also excluded, as the high amplitude of the injected signal caused it to dominate the list of candidates recovered in those bands. In this paper we refer to the candidates with the highest value of the detection statistic as the {\it{loudest}} candidates. 

The highest expected value from Gaussian noise over $N_{\text{trials}}$ independent trials of $2\avF$ is determined\footnote{After a simple change of variable from $2\avF$  to $\Nseg\times 2\avF$.} by numerical integration of the probability density function given, for example, by Eq. 7 of \cite{GalacticCenterSearch}. Fitting to the distribution of the highest $2\avF$ values suggests that $N_{\text{trials}}\simeq\,N_{\text{templ}}$, with $N_{\text{templ}}$ being the number of templates searched.

\begin{figure}[h!]
     \includegraphics[width=\columnwidth]{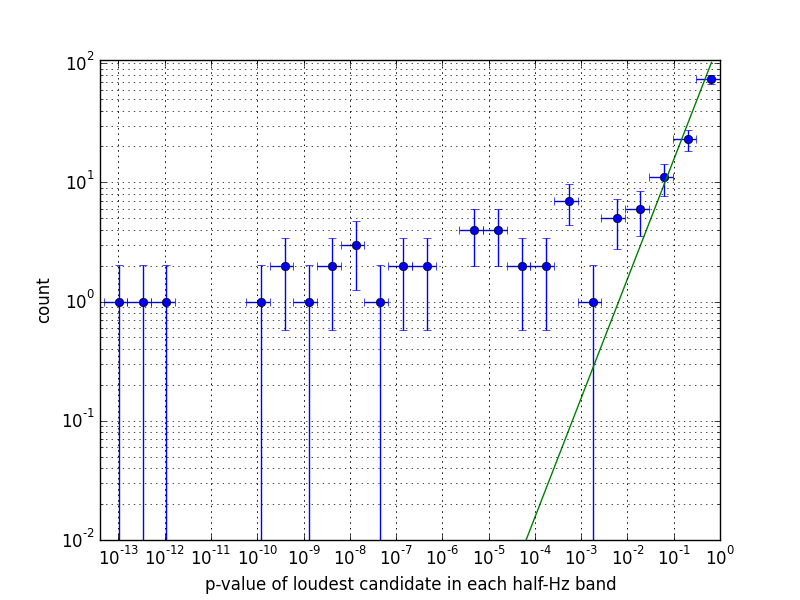}
     \caption{Distribution of p-values, with binomial uncertainties, for the highest detection statistic values measured in half-Hz bands (circles) and expected from pure Gaussian noise (line). We note that the measured p-values for the highest $2\avF$ in the 33.05\,Hz and 35.55\,Hz bands are not shown because they are outside of the x-axis boundaries.}  
\label{fig:pvalues}
\end{figure}

The p-value for the highest $2\avF$ measured in any half-Hz band searched with $N_{\text{trials}}$ independent trials is obtained by integrating the expected noise distribution ($\chi^2_{4\Nseg}$ given in Section \ref{sec:search}) between the observed value and infinity, as done in Eq. 6 of \cite{GalacticCenterSearch}. The distribution of these p-values is shown in Fig. \ref{fig:pvalues} and it is not consistent with what we expect from Gaussian noise across the measured range. Therefore, we can not exclude the presence of a signal in this data based on this distribution alone, as was done in \cite{S6EH}. 

\section{Hierarchical follow Up}
\label{sec:followup}

Since the significance of candidates is not consistent with what we expect from Gaussian noise only, we must investigate ``significant'' candidates to determine if they are produced by a signal or by a detector disturbance. This is done using a hierarchical approach similar to what was used for the hierarchical follow-up of sub-threshold candidates from the Einstein@Home S6 all-sky search \cite{S6EHFU}. 

At each stage of the hierarchical follow-up a semi-coherent search is performed, the top ranking candidates are marked and then searched in the next stage. If the data harbours a real signal, the significance of the recovered candidate will increase with respect to the significance that it had in the previous stage. On the other hand, if the candidate is not produced by a continuous-wave signal, the significance is not expected to increase consistently over the successive stages. 

The hierarchical approach used in this search consists of four stages. This is the smallest number of stages within which we could achieve a fully-coherent search, given the available computing resources. Directly performing a fully-coherent follow-up of all significant candidates from the all-sky search would have been computationally unfeasible.

\begin{table}[t]
\centering
\begin{tabular}{|c|c|c|c|c|c|c|}
\hline
\hline
 & $T_\mathrm{coh}$ & $\Nseg$ & $\delta f$ & $\delta {\dot{f_c}}$ & $\gamma$ & $m_{\text{sky}}$ \\
& hr &  & Hz &  Hz/s & & \\
\hline
\hline
Stage 0 & 210 & 12 & $8.3\ee{-7}$   & $1.3\ee{-11}$ &  100 & $1\ee{-3}$ \\
\hline
Stage 1 & 500 & 5 & $6.7\ee{-7}$   & $2.9\ee{-12}$ &  80 & $8\ee{-6}$ \\
\hline
Stage 2 & 1260 & 2 &  $1.9\ee{-7}$ & $9.3\ee{-13}$& 30  & $1\ee{-6}$\\
\hline
Stage 3 & 2512 & 1 & $6.7\ee{-8}$ & $9.3\ee{-14}$ & 1  & $4\ee{-7}$\\
\hline
\hline
\end{tabular}
\caption{Search parameters for each stage. The follow-up stages are stages 1, 2 and 3. Also shown are the parameters for stage-0, taken from Table \ref{tab:GridSpacings}.}
\label{tab:StagesGridSpacings}
\end{table}

\subsection{Stage 0}
\label{sec:stage0}

We bundle together candidates from the all-sky search that can be ascribed to the same root cause. This clustering step is a standard step in a multi-stage approach \cite{S6EHFU}: Both a loud signal and a loud disturbance produce high values of the detection statistic at a number of different template grid points, and it is a waste of compute cycles to follow up each of these independently. 

\begin{figure}[h!]
     \includegraphics[width=\columnwidth]{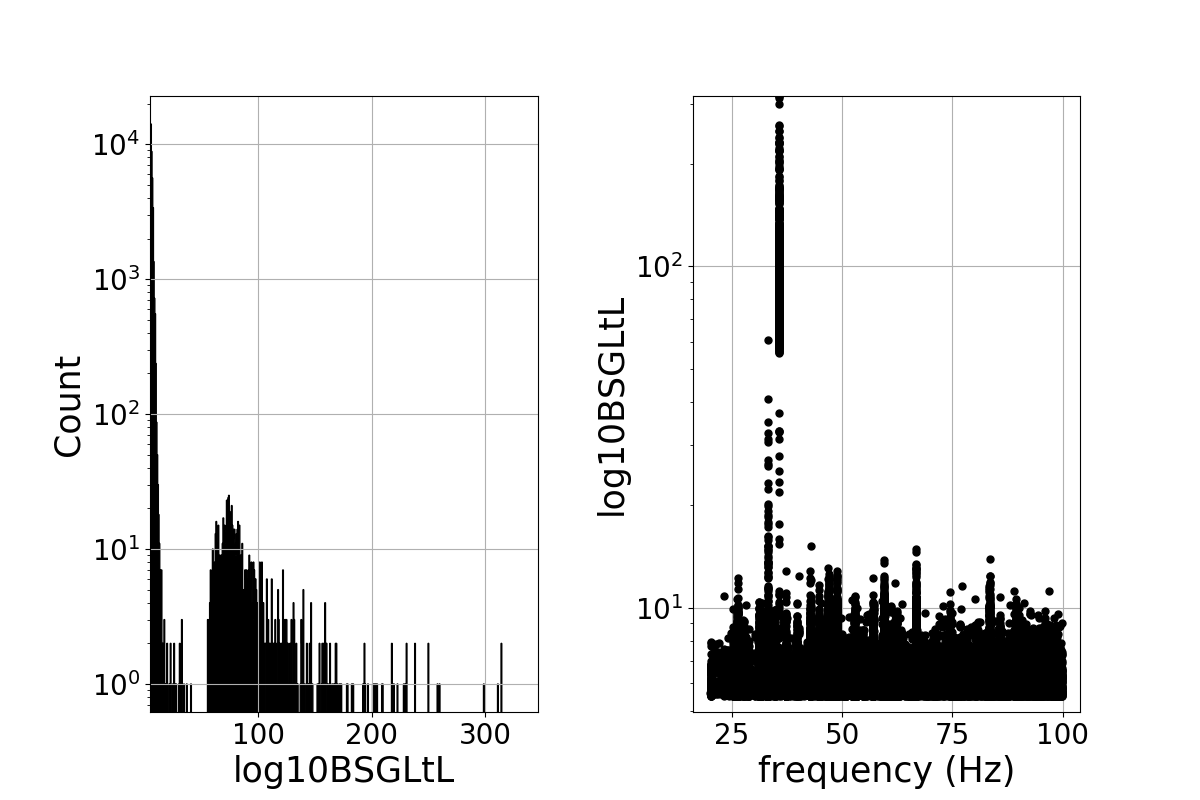}
     \caption{Candidates that are followed-up in stage 1 : the distribution of their detection statistic values $\BSNtsc$ (left plot) and their distribution as a function of frequency (right plot).}
\label{fig:stage0cand}
\end{figure}

We apply a clustering procedure that associates together multiple candidates close to each other in parameter space, and assigns them the parameters of the loudest among them, the seed. We use a new procedure with respect to \cite{S6EHFU} that adapts the cluster size to the data and checks for consistency of the cluster volume with what is expected from a signal \cite{AvAC}. A candidate must have a $\BSNtsc > 5.5$ to be a cluster seed. This threshold is chosen such that only a handful of candidates per 50-mHz would be selected if the data were consistent with Gaussian noise. In this search, there are 15 million candidates with $\BSNtsc > 5.5$. A lower threshold of $\BSNtsc > 4.0$ is applied to candidates that can be included in a cluster. If a cluster has at least two occupants (including the seed), the seed is marked for follow-up. In total, $35963$ seeds are marked for follow up. The $\BSNtsc$ values of these candidates are shown in Fig. \ref{fig:stage0cand} as well as their distribution in frequency.

Monte Carlo studies, using simulated signals added into the data, are conducted to determine how far from the signal parameters a signal candidate is recovered. These signals are simulated at a fixed strain amplitude for which most have $\BSNtsc \gtrapprox 10.0$. We find that $1282$ of $1294$ signal candidates recovered after clustering (99\%) are recovered within 
\begin{equation}
\begin{cases}
 \Delta f &= \pm 9.25\times 10^{-5}\, {\textrm{Hz}}\\
\Delta \dot{f} &= \pm 4.25\times 10^{-11}\, {\textrm{Hz/s}}\\
 \Delta {\text{sky}} &\simeq 4.5 \; {\textrm{sky grid points}}
\end{cases}
\label{eq:FU1region}
\end{equation}
of the signal parameters. This confidence region\footnote{We pick 99\% confidence rather than, say, 100\%, because to reach the 100\% confidence level would require an increase in containment region too large for the available computing resources.} defines the parameter space around each candidate which will be searched in the first stage of the hierarchical follow up. For weaker signals the confidence associated with this uncertainty region decreases. For signals close to the threshold used here, namely with $\BSNtsc$ between 5.5 and 10, the detection confidence only drops by a few percent (see bottom panel of Fig.7 and last row of Table II in \cite{AvAC}).

\subsection{Stage 1}
\label{sec:stage1}

In this stage we search a volume of parameter space (Eqs.~\ref{eq:FU1region}) around each cluster seed. We fix the run time per candidate to be 4 hours on an average CPU of the ATLAS computing cluster \cite{ATLAS}. This yields an optimal search set-up having a coherent baseline of 500 hours, with 5 segments and the grid spacings shown in Table \ref{tab:StagesGridSpacings}. We use the same ranking statistic as the original search, $\BSNtsc$, with tunings updated for $\Nseg = 5$. 

For the population of simulated signals that passed the previous stage, stage 0, $1268$ of $1282$ (99\%) are recovered within the uncertainty region
\begin{equation}
\begin{cases}
 \Delta f &= \pm 1.76\times 10^{-5}\, {\textrm{Hz}}\\
\Delta \dot{f} &= \pm 9.6\times 10^{-12}\, {\textrm{Hz/s}}\\
 \Delta {\text{sky}} &\simeq 0.23~  \Delta {\text{sky}}^{\text{ Stage-0}}.
\end{cases}
\label{eq:FU2region}
\end{equation}

\begin{figure}[h!]
     \includegraphics[width=\columnwidth]{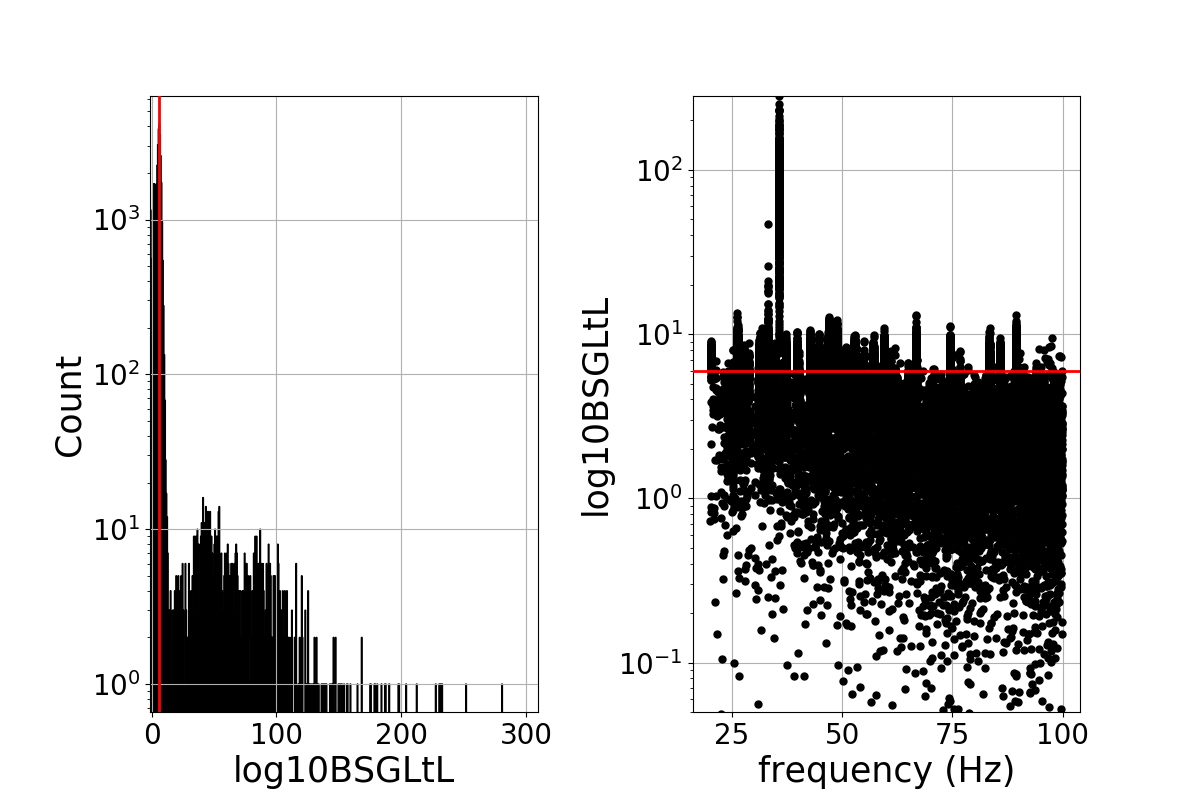}
     \caption{Detection statistic of the loudest candidate from each stage-1 search: the distribution of their detection statistic values $\BSNtsc$ (left plot) and their distribution as a function of frequency (right plot). 411 candidates have $\BSNtsc$ values lower than the axes boundaries on the right plot. The red line marks $\BSNtsc$ = 6.0 which is the threshold at and above which candidates are passed on to stage-2.}
\label{fig:stage1cand}
\end{figure}

From each of the $35963$ follow-up searches we record the most significant candidate in $\BSNtsc$. The distribution of these is shown in Fig. \ref{fig:stage1cand}. A threshold at $\BSNtsc$ = 6.0, derived from Monte Carlo studies, is applied to select the candidates to consider in the next stage. There are $14456$ candidates above this threshold.

\subsection{Stage 2}
\label{sec:stage2}

In this stage we search a volume of parameter space (Eqs.~\ref{eq:FU2region}) around each candidate from stage-1. We fix the run time per candidate to be 4 hours on an average CPU of the ATLAS computing cluster \cite{ATLAS}. This yields an optimal search set-up having a coherent baseline of 1260 hours, with 2 segments and the grid spacings shown in Table \ref{tab:StagesGridSpacings}. We use a different ranking statistic from the original search, because with 2 segments the transient line veto is not useful. Instead we use the ranking statistic $\BSNsc:=\log_{10}\notlogBSNsc$, introduced in~\cite{Keitel:2013} and previously used in \cite{S6EH}, with tunings updated for $\Nseg = 2$. 

For the population of signals that passed the previous stage, $1265$ of $1268$ ($> 99\%$) are recovered within the uncertainty region 
\begin{equation}
\begin{cases}
 \Delta f &= \pm 8.65\times 10^{-6}\, {\textrm{Hz}}\\
\Delta \dot{f} &= \pm 7.8\times 10^{-12}\, {\textrm{Hz/s}}\\
 \Delta {\text{sky}} &\simeq 0.81~  \Delta {\text{sky}}^{\text{ Stage-1}}.
\end{cases}
\label{eq:FU3region}
\end{equation}

\begin{figure}[h!]
     \includegraphics[width=\columnwidth]{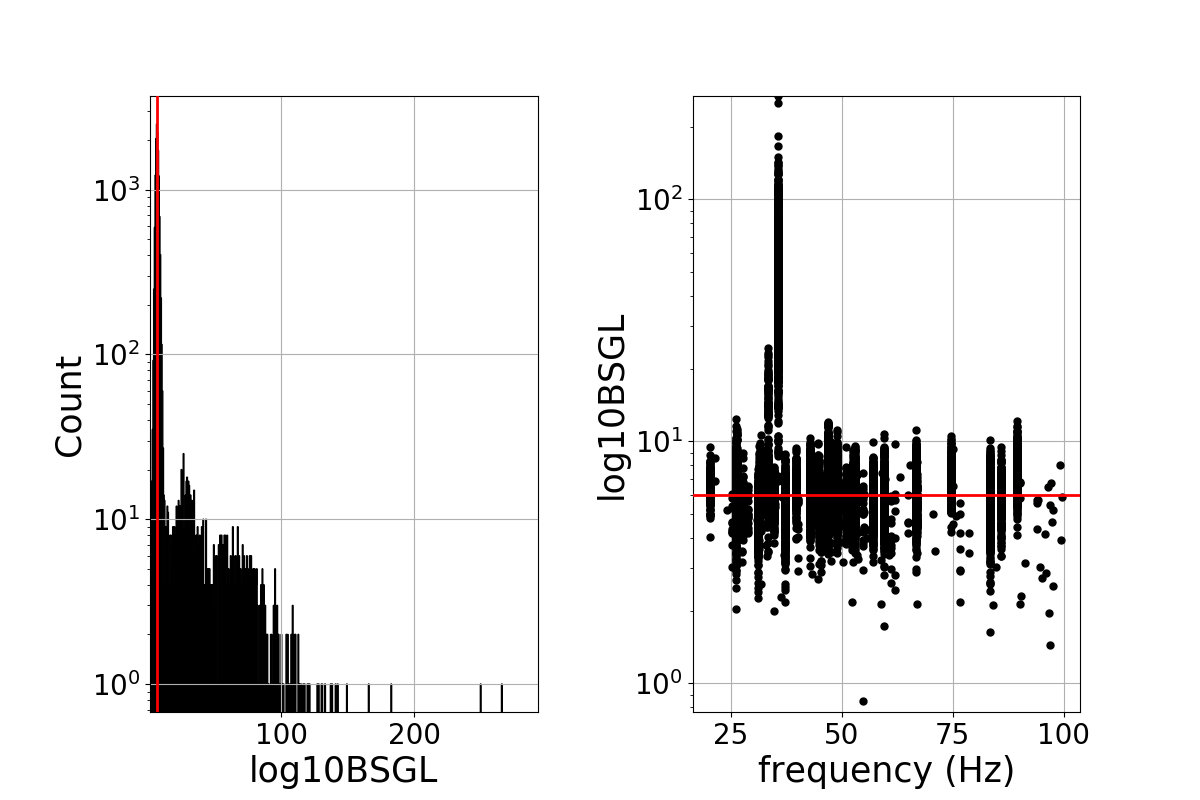}
     \caption{Detection statistic of the loudest candidate from each stage-2 search: the distribution of their detection statistic values $\BSNsc$ (left plot) and their distribution as a function of frequency (right plot). The red line marks $\BSNsc$ = 6.0 which is the threshold at and above which candidates are passed on to stage-3. }
\label{fig:stage2cand}
\end{figure}

From each of the follow-up searches we record the most significant candidate in $\BSNsc$. The distribution of these is shown in Fig. \ref{fig:stage2cand}. A threshold at $\BSNsc$ = 6.0 is applied to determine what candidates to consider in the next stage. There are $8486$ candidates above threshold.

\subsection{Stage 3}
\label{sec:stage3}

In this stage we search a volume of parameter space (Eqs. \ref{eq:FU3region}) around each candidate. We perform a fully coherent search, with a coherent baseline of 2512 hours. The grid spacings are shown in Table \ref{tab:StagesGridSpacings}. We use the same ranking statistic as the previous stage, $\BSNsc$, with tunings updated for $\Nseg = 1$. 

For the population of signals that passed the previous stage, $1265$ of $1265$ ($>99\%$) are recovered within the uncertainty region
\begin{equation}
\begin{cases}
 \Delta f &= \pm 7.5\times 10^{-6}\, {\textrm{Hz}}\\
\Delta \dot{f} &= \pm 7\times 10^{-12}\, {\textrm{Hz/s}}\\
 \Delta {\text{sky}} &\simeq  0.99~ \Delta {\text{sky}}^{\text{ Stage-2}}.
\end{cases}
\label{eq:FU4region}
\end{equation}

This uncertainty region assumes candidates are within the uncertainty regions shown in Eqs.   \ref{eq:FU1region}, \ref{eq:FU2region} and \ref{eq:FU3region} for each of the corresponding follow-up stages. It is possible that a strong candidate which is outside these uncertainty regions would be significant enough to pass through all follow-up stages. In this case the uncertainty on the signal parameters would be larger than the uncertainty region defined in Eq.  \ref{eq:FU4region}.

\begin{figure}[h!]
     \includegraphics[width=\columnwidth]{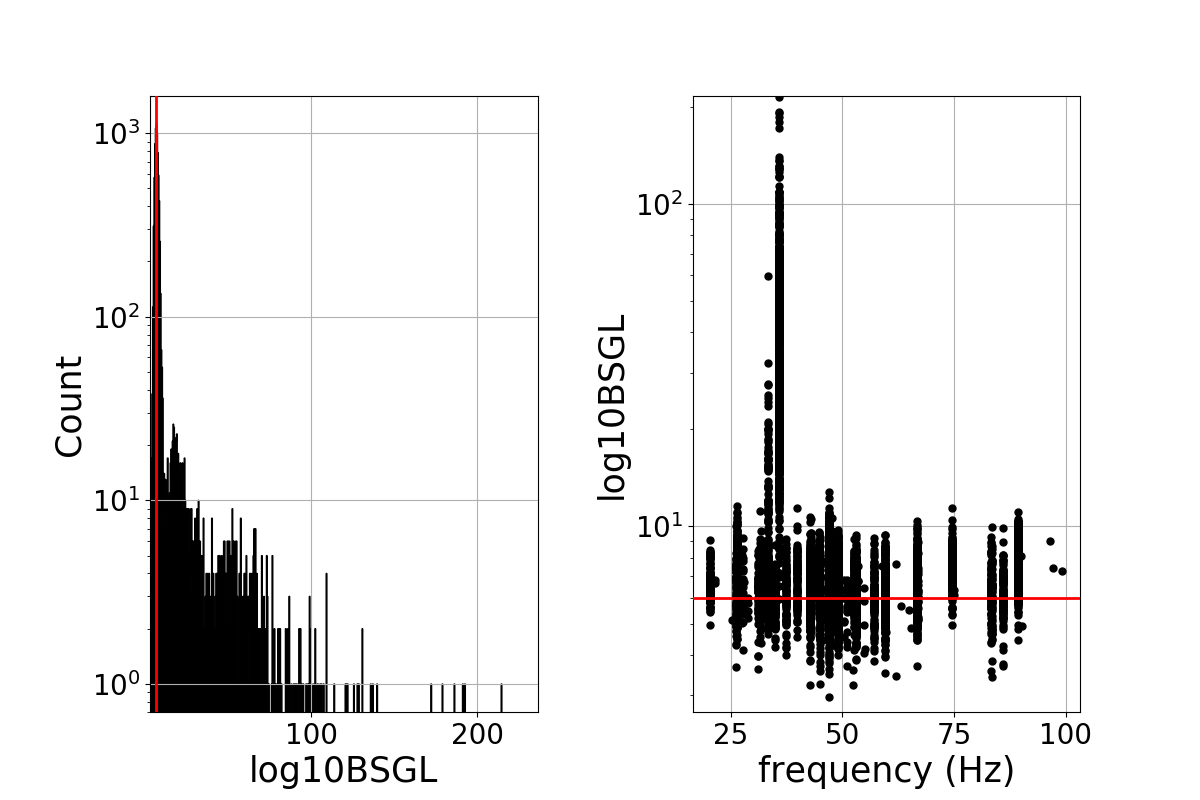}
     \caption{Detection statistic of the loudest candidate from each stage-3 search: the distribution of their detection statistic values $\BSNsc$ (left plot) and their distribution as a function of frequency (right plot). The red line marks $\BSNsc$ = 6.0 which is the threshold below which candidates are discarded.}
\label{fig:stage3cand}
\end{figure}

From each of the follow-up searches we record the most significant candidate in $\BSNsc$. The distribution of these is shown in Fig. \ref{fig:stage3cand}. A threshold at $\BSNsc$ = 6.0 is applied to determine what candidates require further study. There are $6349$ candidates above threshold. Many candidates appear to be from the same feature at a specific frequency. There are 57 distinct narrow frequency regions at which these $6349$ candidates have been recovered.

\subsection{Doppler Modulation off veto}
\label{sec:DM-off}

We employ a newly developed Doppler modulation off (DM-off) veto \cite{DM-off} to determine if the surviving candidates are of terrestrial origin. When searching for CW signals, the frequency of the signal template at any point in time is demodulated for the Doppler effect from the motion of the detectors around the earth and around the sun. If this de-modulation is disabled, a candidate of astrophysical origin would not be recovered with the same significance. In contrast, a candidate of terrestrial origin could potentially become more significant. This is the basis of the DM-off veto. 

For each candidate, the search range of the DM-off searches includes all detector frequencies that could have contributed to the original candidate, accounting for $\dot{f}$ and Doppler corrections. The $\dot{f}$ range includes the original all-sky search range, and extends into large positive values of $\dot{f}$ to allow for a wider range of detector artefact behaviour. 

For a candidate to pass the DM-off veto it must be that its  
$2\F_{\textrm{DM-off}} \leq 2\F^{\textrm{thr}}_{\textrm{DM-off}}$.  
The $2\F^{\textrm{thr}}_{\textrm{DM-off}}$ is picked to be safe, i.e. to not veto any signal candidate with $2\F_\textrm{DM-on}$ in the range of the candidates under consideration. In particular we find that for candidates with $2\F_\textrm{DM-on} < 500$, after the third follow-up, $2\F^{\textrm{thr}}_\textrm{DM-off} = 62$. The threshold increases for candidates with $2\F_{\textrm{DM-on}} > 500$, scaling linearly with the candidates $2\F_{\textrm{DM-on}}$ (see Figure 4 of \cite{DM-off}).

As described in \cite{DM-off}, the DM-off search is first run using data from both detectors and a search grid which is ten times coarser in $f$ and $\dot{f}$ than the stage-3 search. $653$ of the $6349$ candidates pass the $2\F^{\textrm{thr}}_{\textrm{DM-off}}$ threshold. These surviving candidates undergo another similar search, except that the search is performed separately on the data from each of the LIGO detectors. $101$ candidates survive, and undergo a final DM-off search stage. This search uses the fine grid parameters of the stage-3 search (Table \ref{tab:StagesGridSpacings}), covers the parameter space which resulted in the largest $2\F_\textrm{DM-off}$ from the previous DM-off steps, and is performed using both detectors jointly and each detector separately. For a candidate to survive this stage it has to pass {\it {all}} three stage-3 searches.

Four candidates survive the full DM-off veto. The parameters of the candidates, after the third follow-up, are given in Table \ref{tab:LastFourCandidates}. The $2\F_\textrm{DM-off}$ values are also given in this table. 

\begin{table*}[t]
\centering
\begin{tabular}{|c|c|c|c|c|c|c|c|c|c|}
\hline
\hline
ID & $f$  [Hz] & $\alpha$ [rad] & $\delta$ [rad] & ${\dot{f}}$ [Hz/s] & $2\avF$ & $2\avF_{\text{H1}}$ &  $2\avF_{\text{L1}}$ & $2\F_\textrm{DM-off}$\\
\hline
$1 $&$ 58.970435900 $&$ 1.87245 $&$ -0.51971 $&$ -1.081102\times 10^{-9} $&$ 81.4  $&$ 48.5 $&$ 33.4 $&$ 55 $\\
$2 $&$ 62.081409292 $&$ 4.98020 $&$ 0.58542 $&$-2.326246\times 10^{-9} $&$ 81.9 $&$ 45.5 $&$ 39.0 $&$ 52  $\\
$3 $&$ 97.197674733 $&$ 5.88374 $&$-0.76773 $&$2.28614\times 10^{-10} $&$ 86.5  $&$ 55.0 $&$ 31.8 $&$ 58 $\\
$4 $&$ 99.220728369 $&$ 2.842702 $&$-0.469603 $&$-2.498113\times 10^{-9} $&$ 80.2  $&$ 41.4 $&$ 45.8 $&$ 55$\\
\hline
\hline
\end{tabular}
\caption{Stage-3 follow-up results for each of the 4 candidates that survive the DM-off veto. For illustration purposes in the 7th and 8th column we show the values of the average single-detector detection statistics. Typically, for signals, the single-detector values do not exceed the multi-detector $2\avF$. }
\label{tab:LastFourCandidates}
\end{table*}

\subsection{Follow-up in LIGO O2 data}
\label{sec:O2FU}

If the signal candidates surviving the O1 search are standard continuous-wave signals, i.e. continuous wave signals arising from sources that radiate steadily over many years, they should be present in data from the Advanced LIGO's second observing run (O2) with the same parameters. We perform a follow-up search using three months of O2 data, collected from November 30 2016 to February 28 2017.

The candidate parameters in Table \ref{tab:LastFourCandidates} are translated to the O2 midtime, which is the reference time of the new search. The parameter space covered by the search is determined by the uncertainty on the candidate parameters in Eq. \ref{eq:FU3region}. The frequency region is widened to account for the spindown uncertainty. The O2 follow-up covers a frequency range of $\pm 5.15 \times 10^{-4}$ Hz around the candidates.

The search parameters of the O2 follow-up are given in Table \ref{tab:O2Spacings}. The expected loudest $2\avF$ per follow-up search due to Gaussian noise alone, is $52 \pm 3$, assuming independent search templates. 

If a candidate in Table \ref{tab:LastFourCandidates} were due to a signal, the loudest $2\avF$ expected after the follow-up would be the value given in the second column of Table \ref{tab:O2results}. This expected value is obtained by scaling the $2\avF$ in Table \ref{tab:LastFourCandidates} according to the different duration and the different noise levels between the data set used for the third follow-up and the O2 data set. The expected $2\avF$ also folds-in a conservative factor of 0.9 due to a different mismatch of the O2 template grid with respect to the template grid used for the third follow-up. Thus the expected $2\avF$ in Table \ref{tab:O2results} is a conservative estimate for the minimum $2\avF$ that we would expect from a signal candidate. 

The loudest $2\avF$ after the follow-up in O2 data is also given in Table \ref{tab:O2results}. The loudest $2\avF$ recovered for each candidate are $\approx 2\sigma$ below the expected $2\avF$ for a signal candidate. The recovered $2\avF$ are consistent with what is expected from Gaussian data. 
We conclude that it is unlikely that any of the candidates in Table III arises from a long-lived astronomical source of continuous gravitational waves.

\begin{table}[t]
\begin{tabular}{|c|c|}
\hline
\hline
Parameter & Value \\
\hline
\hline
 $\Tcoh$ & 2160 hrs\\
 \hline
 $\Tref$  & 1168447494.5 GPS sec \\
  \hline
$\Nseg$ & 1 \\
  \hline
$\delta f$ & $9.0 \times 10^{-8}$ Hz \\
 \hline
 $\delta {\dot{f_c}}$ & $1.1 \times 10^{-13}$ Hz/s \\
  \hline
$\gamma$ & 1 \\
  \hline
$m_{\text{sky}}$ &$4\ee{-7}$\\
 \hline
\hline
\end{tabular}

\caption{Search parameters, rounded to the first decimal place, for the follow-up of surviving LIGO O1 candidates in LIGO O2 data. $\Tref$ is the reference time that defines the frequency and frequency derivative values.
}
\label{tab:O2Spacings}
\end{table}

\begin{table}[t]
\begin{tabular}{|c|c|c|}
\hline
\hline
 Candidate & Expected $2\avF \pm 1\sigma$  & Loudest $2\avF$ recovered\\
 \hline
1   & $85 \pm 18$ & $44$\\
  \hline
2   & $90 \pm 19$ & $52$\\
  \hline
3   & $84 \pm 18$ & $49$\\
  \hline
4   & $77 \pm 17$ & $47$\\
  \hline
\hline
\end{tabular}

\caption{Highest $2\avF$ expected after the follow-up in O2 data, if the candidates were due to a signal, compared with the highest $2\avF$ recovered from the follow-up. The $2\avF$ expected in Gaussian noise data is $52 \pm 3$.}
\label{tab:O2results}
\end{table}

\section{Results}
\label{sec:results}

\begin{figure*}
   \includegraphics[width=0.8\textwidth]{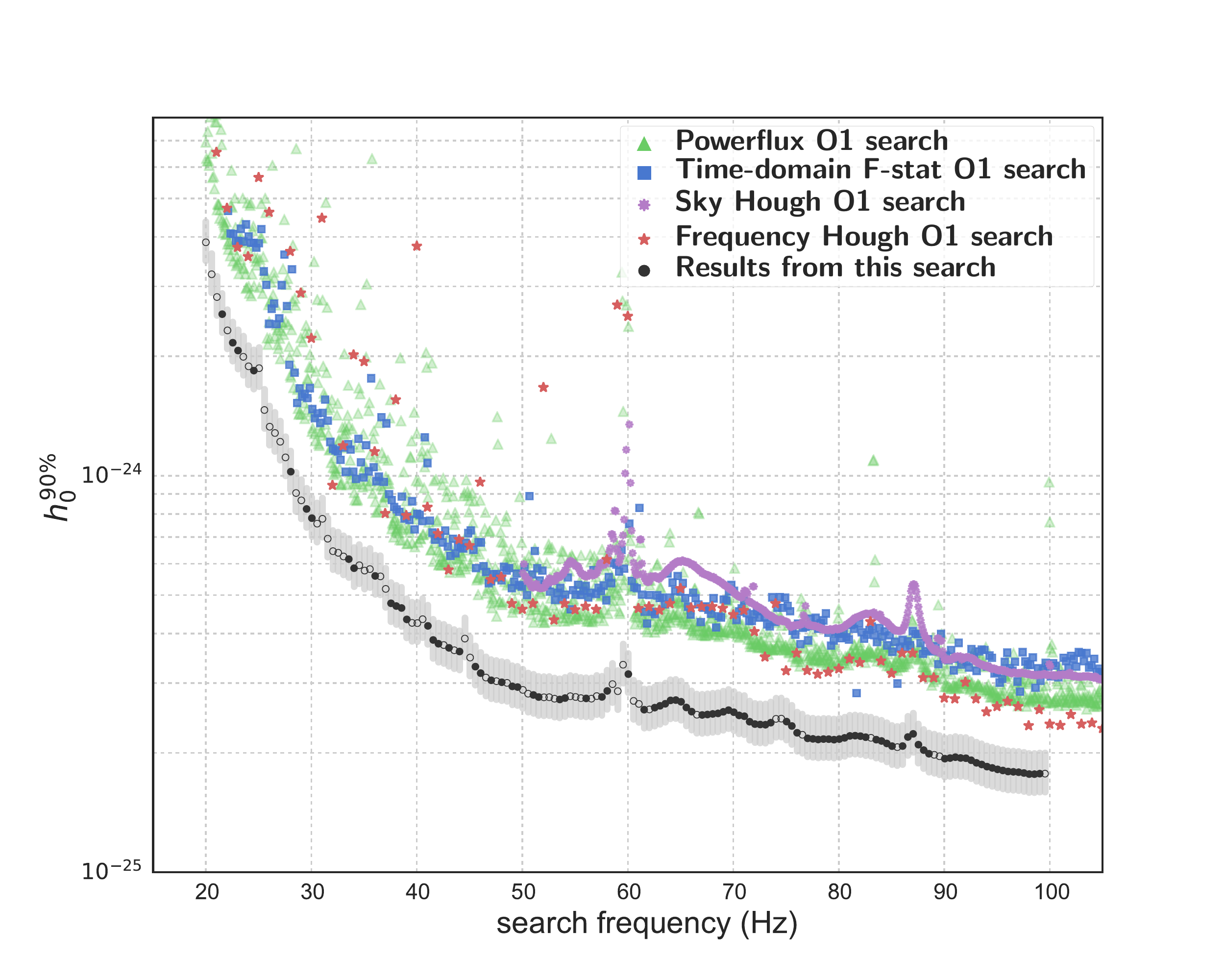}
\caption{90\% confidence upper limits on the gravitational wave amplitude of continuous gravitational wave signals with frequency in 0.5 Hz bands and with spindown values within the searched range. The lowest set of points (black circles) are the results of this search. The empty circles denote half-Hz bands where the upper limit value does not hold for all frequencies in that interval. A list of the excluded frequencies is given in the Appendix. The lighter grey region around the upper limit points shows the 11\% relative difference bracket between upper limits inferred with the procedure described in Section \ref{sec:results} and upper limits that would have been derived (at great computational expense) with direct measurements in all half-Hz bands.  We estimate that less than $\sim 0.5\%$ of the upper limit points would fall outside of this bracket if they were derived with the direct-measurement method in Gaussian noise. For comparison we also plot the most recent upper limits results in this frequency range from O1 data obtained with various search pipelines \cite{O1ASMulti}. We note that these searches cover a broader frequency and spindown range than the search presented here. All upper limits presented here are population-averaged limits over the full sky and source polarisation. 
}  
\label{fig:ULs}
\end{figure*}

The search did not reveal any continuous gravitational wave signal in the parameter volume that was searched. We hence set frequentist 90\% confidence upper limits on the maximum gravitational wave amplitude consistent with this null result in $0.5$ Hz bands,  $h_0^{90\%}(f)$. Specifically, $h_0^{90\%}(f)$ is the GW amplitude such that 90\% of a population of signals with parameter values in our search range would have been detected by our search. We determined the upper limits in bands that were marked as undisturbed in Section \ref{sec:visualInspection}. These upper limits may not hold for frequency bands that were marked as mildly disturbed, which we now consider disturbed as they were excluded by the analysis. These bands, as well as bands which were excluded from further analysis, are identified in Appendix \ref{A:excluded50mHzBands}.

Since an actual full scale fake-signal injection-and-recovery Monte Carlo for the entire set of follow-ups in every $0.5$ Hz band is prohibitive, in the same spirit as \cite{S6EHFU, S6EHCasA, S5GC1HF}, we perform such a study in a limited set of trial bands. We choose 20 half-Hz bands to measure the upper limits. If these half-Hz bands include 50-mHz bands which were not marked undisturbed, no upper limit injections are made in those 50-mHz bands. 

The amplitudes of the fake signals bracket the 90\% confidence region typically between 70\% and 100\%. The $h_0$ versus confidence data is fit in this region with a sigmoid of the form 
\begin{equation}
C(h_0)={1\over{1+\exp({{{\textrm{a}}-h_0}\over{\textrm{b}}})}}
\label{eq:sigmoidFit}
\end{equation}
and the $h_0^{90\%}$ value is read-off of this curve. The fitting procedure\footnote{We used the {\texttt{linfit}} Matlab routine.} yields the best-fit ${\textrm{a}}$ and ${\textrm{b}}$ values and the covariance matrix. Given the binomial confidence values uncertainties, using the covariance matrix we estimate the $h_0^{90\%}$ uncertainty.

For each of these frequency bands we determine the sensitivity depth ${{\mathcal{D}}}^{90\%}$ \cite{GalacticCenterMethod} of the search corresponding to $h_0^{90\%}(f)$: 
\begin{equation}
{{\mathcal{D}}}^{90\%}:={\sqrt{S_h(f)}\over {h_0^{90\%}(f) }}~~[ {1/\sqrt{\text{Hz}}} ],
\label{eq:sensDepth}
\end{equation}
where $\sqrt{S_h(f)}$ is the noise level of the data as a function of frequency.

As representative of the sensitivity depth  of this hierarchical search, we take the average of the measured depths at different frequencies:  $\sensDepth [ {1/\sqrt{\text{Hz}}} ]$. We then determine the 90\% upper limits by substituting this value in Eq.~\ref{eq:sensDepth} for ${{\mathcal{D}}}^{90\%}$.

The upper limit that we get with this procedure, in general yields a different number compared to the upper limit directly measured as done in the twenty test bands.
An $11\%$ relative error bracket comprises the range of variation observed on the measured sensitivity depths, including the uncertainties on the single measurements. So we take this as a generous estimate of the range of variability of the upper limit values introduced by the estimation procedure. 
If the data were Gaussian this bracket would yield a $\sim 0.5\%$ probability of a {\it measured} upper limit falling outside of this bracket. 

Figure \ref{fig:ULs} shows these upper limits as a function of frequency. They are also presented in tabular form in the Appendix with the uncertainties indicating the range of variability introduced by the estimation procedure. The associated uncertainties amount to $\sim 20\%$ when also including 10\% amplitude calibration uncertainty. The most constraining upper limit in the band 98.5-99 Hz, close to the highest frequency, where the detector is most sensitive, is $\lowestUL$. At the lowest end of the frequency range, at 20 Hz, the upper limit rises to $\highestUL$.

\begin{figure}[h!tbp]
   \includegraphics[width=\columnwidth]{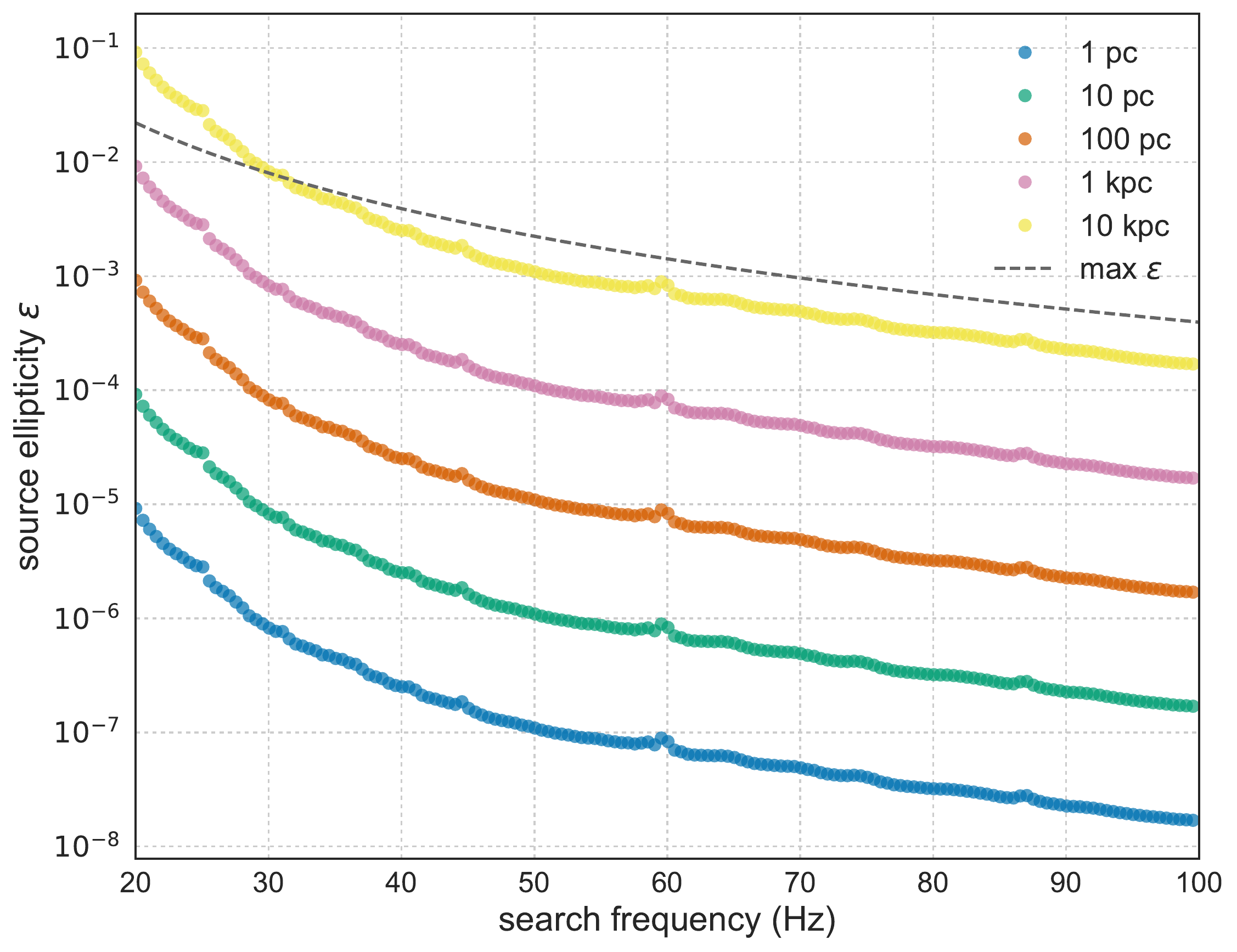}
\caption{Ellipticity $\epsilon$ of a source at a distance d emitting continuous gravitational waves that would have been detected by this search. The dashed line shows the spin-down ellipticity for the highest magnitude spindown parameter value searched: \maxfdot. The spin-down ellipticity is the ellipticity necessary for all the lost rotational kinetic energy to be emitted in gravitational waves. If we assume that the observed spin-down is all actual spin-down of the object, then no ellipticities could be possible above the dashed curve. In reality the observed and actual spindown could differ due to radial acceleration of the source. In this case the actual spin-down of the object may even be larger than the apparent one. In this case our search would be sensitive to objects with ellipticities above the dashed line. }  
\label{fig:reach}
\end{figure}

In general not all the rotational kinetic energy lost is due to GW emission. Following \cite{Ming:2015jla}, we define $x$ to be the fraction of the spindown rotational energy emitted in gravitational waves. The star's ellipticity necessary to sustain such emission is
\begin{equation}
\label{eq:GWspindown}
\epsilon(f,x{\dot{f}})=\sqrt{{5c^5\over 32\pi^4 G}{x{|\dot{f}|}\over If^5}},
\end{equation}
where $c$ is the speed of light, $G$ is the gravitational constant, $f$ is the GW frequency and $I$ the principal moment of inertia of the star. Correspondingly, $x{\dot{f}}$ is the spindown rate that accounts for the emission of GWs and this is why we refer to it as the GW spindown. The gravitational wave amplitude $h_0$ at the detector coming from a GW source like that of Eq. \ref{eq:GWspindown}, at a distance $D$ from Earth is
\begin{equation}
\label{eq:h0}
h_0(f,x\dot{f},D)={1\over D}\sqrt{{5GI\over 2c^3}{x{|\dot{f}|}\over f}}.
\end{equation}
Based on this last equation, we can use the GW amplitude upper limits to bound the minimum distance for compact objects emitting continuous gravitational waves under different assumptions on the object's ellipticity (i.e. gravitational wave spindown). This is shown in Fig. \ref{fig:reach}. Above 55 Hz we can exclude sources with ellipticities larger than $10^{-5}$ within 100 pc of Earth. Rough estimates are that there should be of order $10^4$ neutron stars within this volume.

\section{Conclusions}

This search concentrates the computing power of \EatHs in a relatively small frequency range at low frequencies where all-sky searches are significantly ``cheaper'' than at higher frequencies. For this reason, the initial search could be set-up with a very long coherent observation time of 210 hours and this yields a record sensitivity depth of 48.7 [1/$\sqrt{\textrm{Hz}}$].

The O1 data set in the low frequency range investigated with this search is significantly more polluted by coherent spectral artefacts than most of the data sets from the Initial-LIGO science runs. Because of this, even a relatively high threshold on the detection statistic of the first search yields tens of thousands of candidates, rather than just $O(100)$. We follow each of them up through a hierarchy of three further stages at the end of which $O(7000)$ survive. After the application of a newly developed Doppler-modulation-off veto, 4 survive. These are finally followed up with a fully coherent search using three months of O2 data, which produces results completely consistent with Gaussian noise and falls short of the predictions under the signal hypothesis. We hence proceed to set upper limits on the intrinsic GW amplitude $h_0$. The hierarchical follow-up procedure presented here has also been used to follow-up outliers from other all-sky searches in O1 data with various search pipelines \cite{O1ASMulti}.

The smallest value of the GW amplitude upper limit is $\lowestUL$ in the band 98.5-99 Hz. Fig. \ref{fig:ULs} shows the upper limit values as a function of search frequency. Our upper limits are the tightest ever placed for this population of signals, and are a factor 1.5-2 smaller than the most recent upper limits \cite{O1ASMulti}. We note that \cite{O1ASMulti} presents results from four different all-sky search pipelines covering a broader frequency and spindown range than the one explored here. The coherent time-baseline for all these pipelines is significantly shorter than the 210 hours used by the very first stage of this search. This limits the sensitivity of those searches but it makes them more robust to deviations in the signal waveform from the target waveform, with respect this search.

Translating the upper limits on the GW amplitude in upper limits on the ellipticity of the GW source, we find that for frequencies above 55 Hz our results exclude isolated compact objects with ellipticities of $10^{-5} {\sqrt{10^{38} \textrm{kg\,m}^2/ I}}$ (corresponding to GW spindowns between $10^{-14}$  Hz/s and $10^{-13}$ Hz/s) or higher, within 100 pc of Earth. 

\section{Acknowledgments}

The authors gratefully acknowledge the support of the United States
National Science Foundation (NSF) for the construction and operation of the
LIGO Laboratory and Advanced LIGO as well as the Science and Technology Facilities Council (STFC) of the
United Kingdom, the Max-Planck-Society (MPS), and the State of
Niedersachsen/Germany for support of the construction of Advanced LIGO 
and construction and operation of the GEO600 detector. 
Additional support for Advanced LIGO was provided by the Australian Research Council.
The authors gratefully acknowledge the Italian Istituto Nazionale di Fisica Nucleare (INFN),  
the French Centre National de la Recherche Scientifique (CNRS) and
the Foundation for Fundamental Research on Matter supported by the Netherlands Organisation for Scientific Research, 
for the construction and operation of the Virgo detector
and the creation and support  of the EGO consortium. 
The authors also gratefully acknowledge research support from these agencies as well as by 
the Council of Scientific and Industrial Research of India, 
Department of Science and Technology, India,
Science \& Engineering Research Board (SERB), India,
Ministry of Human Resource Development, India,
the Spanish Ministerio de Econom\'ia y Competitividad,
the  Vicepresid\`encia i Conselleria d'Innovaci\'o, Recerca i Turisme and the Conselleria d'Educaci\'o i Universitat del Govern de les Illes Balears,
the National Science Centre of Poland,
the European Commission,
the Royal Society, 
the Scottish Funding Council, 
the Scottish Universities Physics Alliance, 
the Hungarian Scientific Research Fund (OTKA),
the Lyon Institute of Origins (LIO),
the National Research Foundation of Korea,
Industry Canada and the Province of Ontario through the Ministry of Economic Development and Innovation, 
the Natural Science and Engineering Research Council Canada,
Canadian Institute for Advanced Research,
the Brazilian Ministry of Science, Technology, and Innovation,
International Center for Theoretical Physics South American Institute for Fundamental Research (ICTP-SAIFR), 
Russian Foundation for Basic Research,
the Leverhulme Trust, 
the Research Corporation, 
Ministry of Science and Technology (MOST), Taiwan
and
the Kavli Foundation.
The authors gratefully acknowledge the support of the NSF, STFC, MPS, INFN, CNRS and the
State of Niedersachsen/Germany for provision of computational resources. 

The authors also gratefully acknowledge the support of the many thousands of Einstein@Home volunteers, without whom this search would not have been possible.

This document has been assigned LIGO Laboratory document number \texttt{LIGO-P1700127}.

\newpage


\newpage
\appendix
\section{Tabular data}
\subsection{Upper limit values}
\label{A:ULs}
\onecolumngrid

\begin{longtable}{|c|c|c@{\hskip 0.1in}|c|c|c@{\hskip 0.1in}|c|c|c@{\hskip 0.1in}|c|c|}
\cline{1-2}\cline{4-5}\cline{7-8}\cline{10-11}
$f$ (Hz)& $h_{0}^{90\%}\times 10^{25}$ & & $f$ (Hz) & $h_{0}^{90\%}\times 10^{25}$ & & $f$ (Hz) & $h_{0}^{90\%}\times 10^{25}$ & & $f$ (Hz) & $h_{0}^{90\%}\times 10^{25}$\vspace{0.02in}\\
\cline{1-2}\cline{4-5}\cline{7-8}\cline{10-11}
\endhead
\centering
20.00 &  38.8 $\pm$ 4.9 & & 20.55 &  32.2 $\pm$ 4.1 & & 21.05 &  28.2 $\pm$ 3.6 & & 21.55 &  25.5 $\pm$ 3.3 \\ 
 22.05 &  23.3 $\pm$ 3.0 & & 22.55 &  21.6 $\pm$ 2.8 & & 23.05 &  20.7 $\pm$ 2.6 & & 23.55 &  19.9 $\pm$ 2.5 \\ 
 24.05 &  18.9 $\pm$ 2.4 & & 24.55 &  18.4 $\pm$ 2.3 & & 25.05 &  18.7 $\pm$ 2.4 & & 25.55 &  14.6 $\pm$ 1.9 \\ 
 26.05 &  13.3 $\pm$ 1.7 & & 26.55 &  12.8 $\pm$ 1.6 & & 27.05 &  12.2 $\pm$ 1.6 & & 27.55 &  11.1 $\pm$ 1.4 \\ 
 28.05 &  10.2 $\pm$ 1.3 & & 28.55 &  9.0 $\pm$ 1.2 & & 29.05 &  8.7 $\pm$ 1.1 & & 29.55 &  8.2 $\pm$ 1.0 \\ 
 30.05 &  7.8 $\pm$ 1.0 & & 30.55 &  7.6 $\pm$ 1.0 & & 31.05 &  7.8 $\pm$ 1.0 & & 31.55 &  6.9 $\pm$ 0.9 \\ 
 32.05 &  6.5 $\pm$ 0.8 & & 32.55 &  6.4 $\pm$ 0.8 & & 33.05 &  6.3 $\pm$ 0.8 & & 33.55 &  6.2 $\pm$ 0.8 \\ 
 34.05 &  5.8 $\pm$ 0.7 & & 34.55 &  5.9 $\pm$ 0.8 & & 35.05 &  5.8 $\pm$ 0.7 & & 35.55 &  5.8 $\pm$ 0.7 \\ 
 36.05 &  5.6 $\pm$ 0.7 & & 36.55 &  5.6 $\pm$ 0.7 & & 37.05 &  5.2 $\pm$ 0.7 & & 37.55 &  4.8 $\pm$ 0.6 \\ 
 38.05 &  4.7 $\pm$ 0.6 & & 38.55 &  4.6 $\pm$ 0.6 & & 39.05 &  4.3 $\pm$ 0.6 & & 39.55 &  4.3 $\pm$ 0.5 \\ 
 40.05 &  4.2 $\pm$ 0.5 & & 40.55 &  4.3 $\pm$ 0.6 & & 41.05 &  4.2 $\pm$ 0.5 & & 41.55 &  3.9 $\pm$ 0.5 \\ 
 42.05 &  3.8 $\pm$ 0.5 & & 42.55 &  3.7 $\pm$ 0.5 & & 43.05 &  3.7 $\pm$ 0.5 & & 43.55 &  3.6 $\pm$ 0.5 \\ 
 44.05 &  3.6 $\pm$ 0.5 & & 44.55 &  3.9 $\pm$ 0.5 & & 45.05 &  3.5 $\pm$ 0.4 & & 45.55 &  3.3 $\pm$ 0.4 \\ 
 46.05 &  3.2 $\pm$ 0.4 & & 46.55 &  3.1 $\pm$ 0.4 & & 47.05 &  3.0 $\pm$ 0.4 & & 47.55 &  3.0 $\pm$ 0.4 \\ 
 48.05 &  3.0 $\pm$ 0.4 & & 48.55 &  3.0 $\pm$ 0.4 & & 49.05 &  2.9 $\pm$ 0.4 & & 49.55 &  2.9 $\pm$ 0.4 \\ 
 50.05 &  2.9 $\pm$ 0.4 & & 50.55 &  2.8 $\pm$ 0.4 & & 51.05 &  2.8 $\pm$ 0.4 & & 51.55 &  2.8 $\pm$ 0.4 \\ 
 52.05 &  2.8 $\pm$ 0.4 & & 52.55 &  2.8 $\pm$ 0.4 & & 53.05 &  2.7 $\pm$ 0.3 & & 53.55 &  2.7 $\pm$ 0.3 \\ 
 54.05 &  2.7 $\pm$ 0.3 & & 54.55 &  2.8 $\pm$ 0.4 & & 55.05 &  2.8 $\pm$ 0.4 & & 55.55 &  2.7 $\pm$ 0.3 \\ 
 56.05 &  2.7 $\pm$ 0.3 & & 56.55 &  2.7 $\pm$ 0.3 & & 57.05 &  2.8 $\pm$ 0.4 & & 57.55 &  2.8 $\pm$ 0.4 \\ 
 58.05 &  2.9 $\pm$ 0.4 & & 58.55 &  3.0 $\pm$ 0.4 & & 59.05 &  2.9 $\pm$ 0.4 & & 59.55 &  3.3 $\pm$ 0.4 \\ 
 60.05 &  3.2 $\pm$ 0.4 & & 60.55 &  2.7 $\pm$ 0.3 & & 61.05 &  2.7 $\pm$ 0.3 & & 61.55 &  2.6 $\pm$ 0.3 \\ 
 62.05 &  2.6 $\pm$ 0.3 & & 62.55 &  2.6 $\pm$ 0.3 & & 63.05 &  2.6 $\pm$ 0.3 & & 63.55 &  2.7 $\pm$ 0.3 \\ 
 64.05 &  2.7 $\pm$ 0.3 & & 64.55 &  2.7 $\pm$ 0.3 & & 65.05 &  2.7 $\pm$ 0.3 & & 65.55 &  2.6 $\pm$ 0.3 \\ 
 66.05 &  2.5 $\pm$ 0.3 & & 66.55 &  2.5 $\pm$ 0.3 & & 67.05 &  2.5 $\pm$ 0.3 & & 67.55 &  2.5 $\pm$ 0.3 \\ 
 68.05 &  2.5 $\pm$ 0.3 & & 68.55 &  2.5 $\pm$ 0.3 & & 69.05 &  2.5 $\pm$ 0.3 & & 69.55 &  2.6 $\pm$ 0.3 \\ 
 70.05 &  2.5 $\pm$ 0.3 & & 70.55 &  2.5 $\pm$ 0.3 & & 71.05 &  2.5 $\pm$ 0.3 & & 71.55 &  2.4 $\pm$ 0.3 \\ 
 72.05 &  2.4 $\pm$ 0.3 & & 72.55 &  2.4 $\pm$ 0.3 & & 73.05 &  2.4 $\pm$ 0.3 & & 73.55 &  2.4 $\pm$ 0.3 \\ 
 74.05 &  2.4 $\pm$ 0.3 & & 74.55 &  2.4 $\pm$ 0.3 & & 75.05 &  2.4 $\pm$ 0.3 & & 75.55 &  2.3 $\pm$ 0.3 \\ 
 76.05 &  2.2 $\pm$ 0.3 & & 76.55 &  2.2 $\pm$ 0.3 & & 77.05 &  2.2 $\pm$ 0.3 & & 77.55 &  2.2 $\pm$ 0.3 \\ 
 78.05 &  2.2 $\pm$ 0.3 & & 78.55 &  2.2 $\pm$ 0.3 & & 79.05 &  2.2 $\pm$ 0.3 & & 79.55 &  2.2 $\pm$ 0.3 \\ 
 80.05 &  2.2 $\pm$ 0.3 & & 80.55 &  2.2 $\pm$ 0.3 & & 81.05 &  2.2 $\pm$ 0.3 & & 81.55 &  2.2 $\pm$ 0.3 \\ 
 82.05 &  2.2 $\pm$ 0.3 & & 82.55 &  2.2 $\pm$ 0.3 & & 83.05 &  2.2 $\pm$ 0.3 & & 83.55 &  2.2 $\pm$ 0.3 \\ 
 84.05 &  2.1 $\pm$ 0.3 & & 84.55 &  2.1 $\pm$ 0.3 & & 85.05 &  2.1 $\pm$ 0.3 & & 85.55 &  2.1 $\pm$ 0.3 \\ 
 86.05 &  2.1 $\pm$ 0.3 & & 86.55 &  2.2 $\pm$ 0.3 & & 87.05 &  2.2 $\pm$ 0.3 & & 87.55 &  2.1 $\pm$ 0.3 \\ 
 88.05 &  2.0 $\pm$ 0.3 & & 88.55 &  2.0 $\pm$ 0.3 & & 89.05 &  2.0 $\pm$ 0.3 & & 89.55 &  2.0 $\pm$ 0.2 \\ 
 90.05 &  1.9 $\pm$ 0.2 & & 90.55 &  1.9 $\pm$ 0.2 & & 91.05 &  2.0 $\pm$ 0.2 & & 91.55 &  1.9 $\pm$ 0.2 \\ 
 92.05 &  1.9 $\pm$ 0.2 & & 92.55 &  1.9 $\pm$ 0.2 & & 93.05 &  1.9 $\pm$ 0.2 & & 93.55 &  1.9 $\pm$ 0.2 \\ 
 94.05 &  1.8 $\pm$ 0.2 & & 94.55 &  1.8 $\pm$ 0.2 & & 95.05 &  1.8 $\pm$ 0.2 & & 95.55 &  1.8 $\pm$ 0.2 \\ 
 96.05 &  1.8 $\pm$ 0.2 & & 96.55 &  1.8 $\pm$ 0.2 & & 97.05 &  1.8 $\pm$ 0.2 & & 97.55 &  1.8 $\pm$ 0.2 \\ 
 98.05 &  1.8 $\pm$ 0.2 & & 98.55 &  1.8 $\pm$ 0.2 & & 99.05 &  1.8 $\pm$ 0.2 & & 99.55 &  1.8 $\pm$ 0.2 \\
\cline{1-2}\cline{4-5}\cline{7-8}\cline{10-11}
\caption{First frequency of each half-Hz signal frequency band in which we set upper limits and
upper limit value for that band. The uncertainties correspond to the 11\% relative difference bracket discussed in Section \ref{sec:results}}.
\end{longtable}

\clearpage
\subsection{Cleaned-out frequency bins}
\label{A:cleanedBands}
\twocolumngrid
\begin{longtable}{|cccc|} \hline \hline 
 $f_{\mathrm{L}}$~(Hz)   &    LFS~(Hz)    & HFS~(Hz)           & IFO\\
\hline 
\hline
\endhead
 19.9995 & 0.001 & 0.001 & L \\ 
 20.0 & 0.001 & 0.001 & H \\ 
 20.24999 & 0.001 & 0.001 & H \\ 
 20.25014 & 0.001 & 0.001 & L \\ 
 20.5 & 0.001 & 0.001 & H \\ 
 20.5 & 0.001 & 0.001 & L \\ 
 20.7163 & 0.002 & 0.002 & L \\ 
 20.73 & 0.002 & 0.002 & L \\ 
 20.74121875 & 0.001 & 0.001 & H \\ 
 20.7423125 & 0.001 & 0.001 & H \\ 
 20.9995 & 0.001 & 0.001 & L \\ 
 21.0 & 0.001 & 0.001 & H \\ 
 21.24998 & 0.001 & 0.001 & H \\ 
 21.25011 & 0.001 & 0.001 & L \\ 
 21.3575 & 0.001 & 0.001 & L \\ 
 21.3842 & 0.001 & 0.001 & L \\ 
 21.41043 & 0.001 & 0.001 & L \\ 
 21.41043 & 0.001 & 0.001 & L \\ 
 21.4374 & 0.001 & 0.001 & L \\ 
 21.4639 & 0.001 & 0.001 & L \\ 
 21.499987 & 0.001 & 0.001 & L \\ 
 21.5 & 0.001 & 0.001 & H \\ 
 21.7028 & 0.002 & 0.002 & L \\ 
 21.7165 & 0.002 & 0.002 & L \\ 
 21.7344 & 0.001 & 0.001 & L \\ 
 21.9995 & 0.001 & 0.001 & L \\ 
 22.0 & 0.001 & 0.001 & H \\ 
 22.24997 & 0.001 & 0.001 & H \\ 
 22.25008 & 0.001 & 0.001 & L \\ 
 22.499974 & 0.001 & 0.001 & L \\ 
 22.5 & 0.001 & 0.001 & H \\ 
 22.6893 & 0.002 & 0.002 & L \\ 
 22.7 & 0.0005 & 0.0005 & L \\ 
 22.703 & 0.002 & 0.002 & L \\ 
 22.72233 & 0.001 & 0.001 & L \\ 
 22.815340625 & 0.001 & 0.001 & H \\ 
 22.81654375 & 0.001 & 0.001 & H \\ 
 22.9995 & 0.001 & 0.001 & L \\ 
 23.0 & 0.001 & 0.001 & H \\ 
 23.24996 & 0.001 & 0.001 & H \\ 
 23.25005 & 0.001 & 0.001 & L \\ 
 23.3039 & 0.001 & 0.001 & L \\ 
 23.3306 & 0.001 & 0.001 & L \\ 
 23.35683 & 0.001 & 0.001 & L \\ 
 23.35683 & 0.001 & 0.001 & L \\ 
 23.3838 & 0.001 & 0.001 & L \\ 
 23.4103 & 0.001 & 0.001 & L \\ 
 23.499961 & 0.001 & 0.001 & L \\ 
 23.5 & 0.001 & 0.001 & H \\ 
 23.6758 & 0.002 & 0.002 & L \\ 
 23.6895 & 0.002 & 0.002 & L \\ 
 23.71026 & 0.001 & 0.001 & L \\ 
 23.97079 & 0.0016 & 0.0008 & L \\ 
 23.9995 & 0.001 & 0.001 & L \\ 
 24.0 & 0.0005 & 0.0005 & H \\ 
 24.0 & 0.001 & 0.001 & H \\ 
 24.24995 & 0.001 & 0.001 & H \\ 
 24.25002 & 0.001 & 0.001 & L \\ 
 24.499948 & 0.001 & 0.001 & L \\ 
 24.5 & 0.001 & 0.001 & H \\ 
 24.6623 & 0.002 & 0.002 & L \\ 
 24.676 & 0.002 & 0.002 & L \\ 
 24.69819 & 0.001 & 0.001 & L \\ 
 24.8894625 & 0.001 & 0.001 & H \\ 
 24.890775 & 0.001 & 0.001 & H \\ 
 24.9995 & 0.001 & 0.001 & L \\ 
 25.0 & 0.001 & 0.001 & H \\ 
 25.24994 & 0.001 & 0.001 & H \\ 
 25.24999 & 0.001 & 0.001 & L \\ 
 25.2503 & 0.001 & 0.001 & L \\ 
 25.277 & 0.001 & 0.001 & L \\ 
 25.30323 & 0.001 & 0.001 & L \\ 
 25.30323 & 0.001 & 0.001 & L \\ 
 25.3302 & 0.001 & 0.001 & L \\ 
 25.3567 & 0.001 & 0.001 & L \\ 
 25.499935 & 0.001 & 0.001 & L \\ 
 25.5 & 0.001 & 0.001 & H \\ 
 25.6 & 0.0005 & 0.0005 & L \\ 
 25.6488 & 0.002 & 0.002 & L \\ 
 25.6625 & 0.002 & 0.002 & L \\ 
 25.68612 & 0.001 & 0.001 & L \\ 
 25.9995 & 0.001 & 0.001 & L \\ 
 26.0 & 0.001 & 0.001 & H \\ 
 26.24993 & 0.001 & 0.001 & H \\ 
 26.24996 & 0.001 & 0.001 & L \\ 
 26.499922 & 0.001 & 0.001 & L \\ 
 26.5 & 0.001 & 0.001 & H \\ 
 26.6353 & 0.002 & 0.002 & L \\ 
 26.649 & 0.002 & 0.002 & L \\ 
 26.67405 & 0.001 & 0.001 & L \\ 
 26.963584375 & 0.001 & 0.001 & H \\ 
 26.96500625 & 0.001 & 0.001 & H \\ 
 26.9995 & 0.001 & 0.001 & L \\ 
 27.0 & 0.001 & 0.001 & H \\ 
 27.1967 & 0.001 & 0.001 & L \\ 
 27.2234 & 0.001 & 0.001 & L \\ 
 27.24963 & 0.001 & 0.001 & L \\ 
 27.24963 & 0.001 & 0.001 & L \\ 
 27.24992 & 0.001 & 0.001 & H \\ 
 27.24993 & 0.001 & 0.001 & L \\ 
 27.2766 & 0.001 & 0.001 & L \\ 
 27.3031 & 0.001 & 0.001 & L \\ 
 27.499909 & 0.001 & 0.001 & L \\ 
 27.5 & 0.001 & 0.001 & H \\ 
 27.6218 & 0.002 & 0.002 & L \\ 
 27.6355 & 0.002 & 0.002 & L \\ 
 27.66198 & 0.001 & 0.001 & L \\ 
 27.9995 & 0.001 & 0.001 & L \\ 
 28.0 & 0.001 & 0.001 & H \\ 
 28.2499 & 0.001 & 0.001 & L \\ 
 28.24991 & 0.001 & 0.001 & H \\ 
 28.499896 & 0.001 & 0.001 & L \\ 
 28.5 & 0.001 & 0.001 & H \\ 
 28.5 & 0.0005 & 0.0005 & L \\ 
 28.6083 & 0.002 & 0.002 & L \\ 
 28.622 & 0.002 & 0.002 & L \\ 
 28.64991 & 0.001 & 0.001 & L \\ 
 28.9995 & 0.001 & 0.001 & L \\ 
 29.0 & 0.001 & 0.001 & H \\ 
 29.03770625 & 0.001 & 0.001 & H \\ 
 29.0392375 & 0.001 & 0.001 & H \\ 
 29.1431 & 0.001 & 0.001 & L \\ 
 29.1698 & 0.001 & 0.001 & L \\ 
 29.19603 & 0.001 & 0.001 & L \\ 
 29.19603 & 0.001 & 0.001 & L \\ 
 29.223 & 0.001 & 0.001 & L \\ 
 29.2495 & 0.001 & 0.001 & L \\ 
 29.24987 & 0.001 & 0.001 & L \\ 
 29.2499 & 0.001 & 0.001 & H \\ 
 29.2767 & 0.001 & 0.001 & L \\ 
 29.3031 & 0.001 & 0.001 & L \\ 
 29.499883 & 0.001 & 0.001 & L \\ 
 29.5 & 0.001 & 0.001 & H \\ 
 29.5948 & 0.002 & 0.002 & L \\ 
 29.6085 & 0.002 & 0.002 & L \\ 
 29.63784 & 0.001 & 0.001 & L \\ 
 29.9995 & 0.001 & 0.001 & L \\ 
 30.0 & 0.001 & 0.001 & H \\ 
 30.24984 & 0.001 & 0.001 & L \\ 
 30.24989 & 0.001 & 0.001 & H \\ 
 30.49987 & 0.001 & 0.001 & L \\ 
 30.5 & 0.001 & 0.001 & H \\ 
 30.5813 & 0.002 & 0.002 & L \\ 
 30.595 & 0.002 & 0.002 & L \\ 
 30.62577 & 0.001 & 0.001 & L \\ 
 30.943 & 0.001 & 0.001 & H \\ 
 30.9738 & 0.001 & 0.001 & H \\ 
 30.9995 & 0.001 & 0.001 & L \\ 
 31.0 & 0.001 & 0.001 & H \\ 
 31.0895 & 0.001 & 0.001 & L \\ 
 31.111828125 & 0.001 & 0.001 & H \\ 
 31.11346875 & 0.001 & 0.001 & H \\ 
 31.1162 & 0.001 & 0.001 & L \\ 
 31.14243 & 0.001 & 0.001 & L \\ 
 31.14243 & 0.001 & 0.001 & L \\ 
 31.1694 & 0.001 & 0.001 & L \\ 
 31.1959 & 0.001 & 0.001 & L \\ 
 31.2231 & 0.001 & 0.001 & L \\ 
 31.2495 & 0.001 & 0.001 & L \\ 
 31.24981 & 0.001 & 0.001 & L \\ 
 31.24988 & 0.001 & 0.001 & H \\ 
 31.4 & 0.0005 & 0.0005 & L \\ 
 31.4127 & 0.003 & 0.003 & H \\ 
 31.4149 & 0.003 & 0.003 & H \\ 
 31.499857 & 0.001 & 0.001 & L \\ 
 31.5 & 0.001 & 0.001 & H \\ 
 31.5678 & 0.002 & 0.002 & L \\ 
 31.5815 & 0.002 & 0.002 & L \\ 
 31.6137 & 0.001 & 0.001 & L \\ 
 31.94116 & 0.001 & 0.001 & H \\ 
 31.973 & 0.001 & 0.001 & H \\ 
 31.9995 & 0.001 & 0.001 & L \\ 
 32.0 & 0.0005 & 0.0005 & H \\ 
 32.0 & 0.001 & 0.001 & H \\ 
 32.24978 & 0.001 & 0.001 & L \\ 
 32.24987 & 0.001 & 0.001 & H \\ 
 32.499844 & 0.001 & 0.001 & L \\ 
 32.5 & 0.001 & 0.001 & H \\ 
 33.7 & 0.01556 & 0.01556 & L \\ 
 33.8 & 0.0005 & 0.0005 & L \\ 
 34.3 & 0.0005 & 0.0005 & L \\ 
 34.7 & 0.02778 & 0.02778 & H \\ 
 34.7 & 0.13 & 0.13 & L \\ 
 35.3 & 0.02778 & 0.02778 & H \\ 
 35.3 & 0.13 & 0.13 & L \\ 
 35.706385 & 0.003055 & 0.003055 & L \\ 
 35.7095265 & 0.01222 & 0.01222 & H \\ 
 35.9 & 0.10222 & 0.10222 & H \\ 
 35.958055 & 0.009165 & 0.009165 & L \\ 
 36.7 & 0.10722 & 0.10722 & H \\ 
 36.7 & 0.0005 & 0.0005 & L \\ 
 37.3 & 0.01 & 0.01 & H \\ 
 38.955 & 0.001 & 0.001 & L \\ 
 38.9674 & 0.001 & 0.001 & H \\ 
 38.9815 & 0.001 & 0.001 & L \\ 
 38.9995 & 0.001 & 0.001 & L \\ 
 39.0 & 0.001 & 0.001 & H \\ 
 39.0087 & 0.001 & 0.001 & L \\ 
 39.0351 & 0.001 & 0.001 & L \\ 
 39.24957 & 0.001 & 0.001 & L \\ 
 39.2498 & 0.001 & 0.001 & H \\ 
 39.408315625 & 0.001 & 0.001 & H \\ 
 39.41039375 & 0.001 & 0.001 & H \\ 
 39.4598 & 0.002 & 0.002 & L \\ 
 39.4735 & 0.002 & 0.002 & L \\ 
 39.499753 & 0.001 & 0.001 & L \\ 
 39.5 & 0.001 & 0.001 & H \\ 
 39.51714 & 0.001 & 0.001 & L \\ 
 39.6 & 0.0005 & 0.0005 & L \\ 
 39.92644 & 0.001 & 0.001 & H \\ 
 39.9666 & 0.001 & 0.001 & H \\ 
 39.9995 & 0.001 & 0.001 & L \\ 
 40.0 & 0.0005 & 0.0005 & H \\ 
 40.0 & 0.001 & 0.001 & H \\ 
 40.24954 & 0.001 & 0.001 & L \\ 
 40.24979 & 0.001 & 0.001 & H \\ 
 40.4463 & 0.002 & 0.002 & L \\ 
 40.46 & 0.002 & 0.002 & L \\ 
 40.49974 & 0.001 & 0.001 & L \\ 
 40.5 & 0.001 & 0.001 & H \\ 
 40.50507 & 0.001 & 0.001 & L \\ 
 40.8215 & 0.001 & 0.001 & L \\ 
 40.8482 & 0.001 & 0.001 & L \\ 
 40.87443 & 0.001 & 0.001 & L \\ 
 40.87443 & 0.001 & 0.001 & L \\ 
 40.9014 & 0.001 & 0.001 & L \\ 
 40.9246 & 0.001 & 0.001 & H \\ 
 40.9279 & 0.001 & 0.001 & L \\ 
 40.9551 & 0.001 & 0.001 & L \\ 
 40.9658 & 0.001 & 0.001 & H \\ 
 40.9815 & 0.001 & 0.001 & L \\ 
 40.9995 & 0.001 & 0.001 & L \\ 
 41.0 & 0.001 & 0.001 & H \\ 
 41.24951 & 0.001 & 0.001 & L \\ 
 41.24978 & 0.001 & 0.001 & H \\ 
 41.4328 & 0.002 & 0.002 & L \\ 
 41.4465 & 0.002 & 0.002 & L \\ 
 41.4824375 & 0.001 & 0.001 & H \\ 
 41.484625 & 0.001 & 0.001 & H \\ 
 41.493 & 0.001 & 0.001 & L \\ 
 41.499727 & 0.001 & 0.001 & L \\ 
 41.5 & 0.001 & 0.001 & H \\ 
 41.92276 & 0.001 & 0.001 & H \\ 
 41.965 & 0.001 & 0.001 & H \\ 
 41.9995 & 0.001 & 0.001 & L \\ 
 42.0 & 0.001 & 0.001 & H \\ 
 42.24948 & 0.001 & 0.001 & L \\ 
 42.24977 & 0.001 & 0.001 & H \\ 
 42.4193 & 0.002 & 0.002 & L \\ 
 42.433 & 0.002 & 0.002 & L \\ 
 42.48093 & 0.001 & 0.001 & L \\ 
 42.499714 & 0.001 & 0.001 & L \\ 
 42.5 & 0.001 & 0.001 & H \\ 
 42.5 & 0.0005 & 0.0005 & L \\ 
 42.7679 & 0.001 & 0.001 & L \\ 
 42.7946 & 0.001 & 0.001 & L \\ 
 42.82083 & 0.001 & 0.001 & L \\ 
 42.82083 & 0.001 & 0.001 & L \\ 
 42.8478 & 0.001 & 0.001 & L \\ 
 42.8743 & 0.001 & 0.001 & L \\ 
 42.9015 & 0.001 & 0.001 & L \\ 
 42.92092 & 0.001 & 0.001 & H \\ 
 42.9279 & 0.001 & 0.001 & L \\ 
 42.9642 & 0.001 & 0.001 & H \\ 
 42.9995 & 0.001 & 0.001 & L \\ 
 43.0 & 0.001 & 0.001 & H \\ 
 43.24945 & 0.001 & 0.001 & L \\ 
 43.24976 & 0.001 & 0.001 & H \\ 
 43.4058 & 0.002 & 0.002 & L \\ 
 43.4195 & 0.002 & 0.002 & L \\ 
 43.46886 & 0.001 & 0.001 & L \\ 
 43.499701 & 0.001 & 0.001 & L \\ 
 43.5 & 0.001 & 0.001 & H \\ 
 43.556559375 & 0.001 & 0.001 & H \\ 
 43.55885625 & 0.001 & 0.001 & H \\ 
 43.91908 & 0.001 & 0.001 & H \\ 
 43.9634 & 0.001 & 0.001 & H \\ 
 43.9995 & 0.001 & 0.001 & L \\ 
 44.0 & 0.001 & 0.001 & H \\ 
 44.24942 & 0.001 & 0.001 & L \\ 
 44.24975 & 0.001 & 0.001 & H \\ 
 44.3923 & 0.002 & 0.002 & L \\ 
 44.406 & 0.002 & 0.002 & L \\ 
 44.45679 & 0.001 & 0.001 & L \\ 
 44.499688 & 0.001 & 0.001 & L \\ 
 44.5 & 0.001 & 0.001 & H \\ 
 44.7143 & 0.001 & 0.001 & L \\ 
 44.741 & 0.001 & 0.001 & L \\ 
 44.76723 & 0.001 & 0.001 & L \\ 
 44.76723 & 0.001 & 0.001 & L \\ 
 44.7942 & 0.001 & 0.001 & L \\ 
 44.8207 & 0.001 & 0.001 & L \\ 
 44.8479 & 0.001 & 0.001 & L \\ 
 44.8743 & 0.001 & 0.001 & L \\ 
 44.91724 & 0.001 & 0.001 & H \\ 
 44.9626 & 0.001 & 0.001 & H \\ 
 44.9995 & 0.001 & 0.001 & L \\ 
 45.0 & 0.001 & 0.001 & H \\ 
 45.24939 & 0.001 & 0.001 & L \\ 
 45.24974 & 0.001 & 0.001 & H \\ 
 45.3788 & 0.002 & 0.002 & L \\ 
 45.3925 & 0.002 & 0.002 & L \\ 
 45.4 & 0.0005 & 0.0005 & L \\ 
 45.44472 & 0.001 & 0.001 & L \\ 
 45.499675 & 0.001 & 0.001 & L \\ 
 45.5 & 0.001 & 0.001 & H \\ 
 45.63068125 & 0.001 & 0.001 & H \\ 
 45.6330875 & 0.001 & 0.001 & H \\ 
 45.9 & 0.0005 & 0.0005 & L \\ 
 45.9154 & 0.001 & 0.001 & H \\ 
 45.9618 & 0.001 & 0.001 & H \\ 
 45.9995 & 0.001 & 0.001 & L \\ 
 46.0 & 0.001 & 0.001 & H \\ 
 46.24936 & 0.001 & 0.001 & L \\ 
 46.24973 & 0.001 & 0.001 & H \\ 
 46.3653 & 0.002 & 0.002 & L \\ 
 46.379 & 0.002 & 0.002 & L \\ 
 46.43265 & 0.001 & 0.001 & L \\ 
 46.499662 & 0.001 & 0.001 & L \\ 
 46.5 & 0.001 & 0.001 & H \\ 
 46.6607 & 0.001 & 0.001 & L \\ 
 46.6874 & 0.001 & 0.001 & L \\ 
 46.71363 & 0.001 & 0.001 & L \\ 
 46.71363 & 0.001 & 0.001 & L \\ 
 46.7406 & 0.001 & 0.001 & L \\ 
 46.7671 & 0.001 & 0.001 & L \\ 
 46.7943 & 0.001 & 0.001 & L \\ 
 46.8207 & 0.001 & 0.001 & L \\ 
 46.91356 & 0.001 & 0.001 & H \\ 
 46.961 & 0.001 & 0.001 & H \\ 
 46.9995 & 0.001 & 0.001 & L \\ 
 47.0 & 0.001 & 0.001 & H \\ 
 47.24933 & 0.001 & 0.001 & L \\ 
 47.24972 & 0.001 & 0.001 & H \\ 
 47.3518 & 0.002 & 0.002 & L \\ 
 47.3655 & 0.002 & 0.002 & L \\ 
 47.42058 & 0.001 & 0.001 & L \\ 
 47.499649 & 0.001 & 0.001 & L \\ 
 47.5 & 0.001 & 0.001 & H \\ 
 47.704803125 & 0.001 & 0.001 & H \\ 
 47.70731875 & 0.001 & 0.001 & H \\ 
 47.8 & 0.0005 & 0.0005 & L \\ 
 47.91172 & 0.001 & 0.001 & H \\ 
 47.94158 & 0.0032 & 0.0016 & L \\ 
 47.9602 & 0.001 & 0.001 & H \\ 
 47.9995 & 0.001 & 0.001 & L \\ 
 48.0 & 0.0005 & 0.0005 & H \\ 
 48.0 & 0.001 & 0.001 & H \\ 
 48.2493 & 0.001 & 0.001 & L \\ 
 48.24971 & 0.001 & 0.001 & H \\ 
 48.3 & 0.0005 & 0.0005 & L \\ 
 48.3383 & 0.002 & 0.002 & L \\ 
 48.352 & 0.002 & 0.002 & L \\ 
 48.40851 & 0.001 & 0.001 & L \\ 
 48.499636 & 0.001 & 0.001 & L \\ 
 48.5 & 0.001 & 0.001 & H \\ 
 48.6071 & 0.001 & 0.001 & L \\ 
 48.6338 & 0.001 & 0.001 & L \\ 
 48.66003 & 0.001 & 0.001 & L \\ 
 48.66003 & 0.001 & 0.001 & L \\ 
 48.687 & 0.001 & 0.001 & L \\ 
 48.7135 & 0.001 & 0.001 & L \\ 
 48.7407 & 0.001 & 0.001 & L \\ 
 48.7671 & 0.001 & 0.001 & L \\ 
 48.90988 & 0.001 & 0.001 & H \\ 
 48.9594 & 0.001 & 0.001 & H \\ 
 48.9995 & 0.001 & 0.001 & L \\ 
 49.0 & 0.001 & 0.001 & H \\ 
 49.24927 & 0.001 & 0.001 & L \\ 
 49.2497 & 0.001 & 0.001 & H \\ 
 49.3248 & 0.002 & 0.002 & L \\ 
 49.3385 & 0.002 & 0.002 & L \\ 
 49.499623 & 0.001 & 0.001 & L \\ 
 49.5 & 0.001 & 0.001 & H \\ 
 49.778925 & 0.001 & 0.001 & H \\ 
 49.78155 & 0.001 & 0.001 & H \\ 
 49.90804 & 0.001 & 0.001 & H \\ 
 49.9995 & 0.001 & 0.001 & L \\ 
 50.0 & 0.001 & 0.001 & H \\ 
 50.24924 & 0.001 & 0.001 & L \\ 
 50.3113 & 0.002 & 0.002 & L \\ 
 50.325 & 0.002 & 0.002 & L \\ 
 50.49961 & 0.001 & 0.001 & L \\ 
 50.5 & 0.001 & 0.001 & H \\ 
 50.5535 & 0.001 & 0.001 & L \\ 
 50.5802 & 0.001 & 0.001 & L \\ 
 50.60643 & 0.001 & 0.001 & L \\ 
 50.60643 & 0.001 & 0.001 & L \\ 
 50.6334 & 0.001 & 0.001 & L \\ 
 50.6599 & 0.001 & 0.001 & L \\ 
 50.6871 & 0.001 & 0.001 & L \\ 
 50.7135 & 0.001 & 0.001 & L \\ 
 50.9062 & 0.001 & 0.001 & H \\ 
 51.0 & 0.001 & 0.001 & H \\ 
 51.2 & 0.0005 & 0.0005 & L \\ 
 51.24921 & 0.001 & 0.001 & L \\ 
 51.2978 & 0.002 & 0.002 & L \\ 
 51.3115 & 0.002 & 0.002 & L \\ 
 51.499597 & 0.001 & 0.001 & L \\ 
 51.5 & 0.001 & 0.001 & H \\ 
 51.853046875 & 0.001 & 0.001 & H \\ 
 51.85578125 & 0.001 & 0.001 & H \\ 
 51.90436 & 0.001 & 0.001 & H \\ 
 52.0 & 0.001 & 0.001 & H \\ 
 52.24918 & 0.001 & 0.001 & L \\ 
 52.2843 & 0.002 & 0.002 & L \\ 
 52.298 & 0.002 & 0.002 & L \\ 
 52.499584 & 0.001 & 0.001 & L \\ 
 52.4999 & 0.001 & 0.001 & L \\ 
 52.5 & 0.001 & 0.001 & H \\ 
 52.5266 & 0.001 & 0.001 & L \\ 
 52.55283 & 0.001 & 0.001 & L \\ 
 52.55283 & 0.001 & 0.001 & L \\ 
 52.5798 & 0.001 & 0.001 & L \\ 
 52.6063 & 0.001 & 0.001 & L \\ 
 52.6335 & 0.001 & 0.001 & L \\ 
 52.6599 & 0.001 & 0.001 & L \\ 
 52.90252 & 0.001 & 0.001 & H \\ 
 53.0 & 0.001 & 0.001 & H \\ 
 53.24915 & 0.001 & 0.001 & L \\ 
 53.2708 & 0.002 & 0.002 & L \\ 
 53.2845 & 0.002 & 0.002 & L \\ 
 53.499571 & 0.001 & 0.001 & L \\ 
 53.5 & 0.001 & 0.001 & H \\ 
 53.90068 & 0.001 & 0.001 & H \\ 
 53.92716875 & 0.001 & 0.001 & H \\ 
 53.9300125 & 0.001 & 0.001 & H \\ 
 54.0 & 0.001 & 0.001 & H \\ 
 54.1 & 0.0005 & 0.0005 & L \\ 
 54.2573 & 0.002 & 0.002 & L \\ 
 54.271 & 0.002 & 0.002 & L \\ 
 54.4463 & 0.001 & 0.001 & L \\ 
 54.473 & 0.001 & 0.001 & L \\ 
 54.49923 & 0.001 & 0.001 & L \\ 
 54.49923 & 0.001 & 0.001 & L \\ 
 54.499558 & 0.001 & 0.001 & L \\ 
 54.5 & 0.001 & 0.001 & H \\ 
 54.5262 & 0.001 & 0.001 & L \\ 
 54.5527 & 0.001 & 0.001 & L \\ 
 54.5799 & 0.001 & 0.001 & L \\ 
 54.6063 & 0.001 & 0.001 & L \\ 
 54.89884 & 0.001 & 0.001 & H \\ 
 55.0 & 0.001 & 0.001 & H \\ 
 55.2438 & 0.002 & 0.002 & L \\ 
 55.2575 & 0.002 & 0.002 & L \\ 
 55.499545 & 0.001 & 0.001 & L \\ 
 55.5 & 0.001 & 0.001 & H \\ 
 55.897 & 0.001 & 0.001 & H \\ 
 56.0 & 0.0005 & 0.0005 & H \\ 
 56.0 & 0.001 & 0.001 & H \\ 
 56.001290625 & 0.001 & 0.001 & H \\ 
 56.00424375 & 0.001 & 0.001 & H \\ 
 56.3927 & 0.001 & 0.001 & L \\ 
 56.4194 & 0.001 & 0.001 & L \\ 
 56.44563 & 0.001 & 0.001 & L \\ 
 56.44563 & 0.001 & 0.001 & L \\ 
 56.4726 & 0.001 & 0.001 & L \\ 
 56.4991 & 0.001 & 0.001 & L \\ 
 56.499532 & 0.001 & 0.001 & L \\ 
 56.5 & 0.001 & 0.001 & H \\ 
 56.5 & 0.0005 & 0.0005 & L \\ 
 56.5263 & 0.001 & 0.001 & L \\ 
 56.5527 & 0.001 & 0.001 & L \\ 
 56.89516 & 0.001 & 0.001 & H \\ 
 57.0 & 0.001 & 0.001 & H \\ 
 57.0 & 0.0005 & 0.0005 & L \\ 
 57.499519 & 0.001 & 0.001 & L \\ 
 57.5 & 0.001 & 0.001 & H \\ 
 57.89332 & 0.001 & 0.001 & H \\ 
 58.0 & 0.001 & 0.001 & H \\ 
 58.0754125 & 0.001 & 0.001 & H \\ 
 58.078475 & 0.001 & 0.001 & H \\ 
 58.3391 & 0.001 & 0.001 & L \\ 
 58.3658 & 0.001 & 0.001 & L \\ 
 58.39203 & 0.001 & 0.001 & L \\ 
 58.39203 & 0.001 & 0.001 & L \\ 
 58.419 & 0.001 & 0.001 & L \\ 
 58.4455 & 0.001 & 0.001 & L \\ 
 58.499506 & 0.001 & 0.001 & L \\ 
 58.5 & 0.001 & 0.001 & H \\ 
 58.89148 & 0.001 & 0.001 & H \\ 
 59.0 & 0.001 & 0.001 & H \\ 
 59.499493 & 0.001 & 0.001 & L \\ 
 59.5 & 0.001 & 0.001 & H \\ 
 59.88964 & 0.001 & 0.001 & H \\ 
 59.926975 & 0.004 & 0.002 & L \\ 
 60.0 & 0.001 & 0.001 & H \\ 
 60.0 & 0.06 & 0.06 & H \\ 
 60.0 & 0.06 & 0.06 & L \\ 
 60.149534375 & 0.001 & 0.001 & H \\ 
 60.15270625 & 0.001 & 0.001 & H \\ 
 60.2855 & 0.001 & 0.001 & L \\ 
 60.3122 & 0.001 & 0.001 & L \\ 
 60.33843 & 0.001 & 0.001 & L \\ 
 60.33843 & 0.001 & 0.001 & L \\ 
 60.3654 & 0.001 & 0.001 & L \\ 
 60.3919 & 0.001 & 0.001 & L \\ 
 60.49948 & 0.001 & 0.001 & L \\ 
 60.5 & 0.001 & 0.001 & H \\ 
 60.8878 & 0.001 & 0.001 & H \\ 
 61.0 & 0.001 & 0.001 & H \\ 
 61.499467 & 0.001 & 0.001 & L \\ 
 61.5 & 0.001 & 0.001 & H \\ 
 62.0 & 0.001 & 0.001 & H \\ 
 62.22365625 & 0.001 & 0.001 & H \\ 
 62.2269375 & 0.001 & 0.001 & H \\ 
 62.28483 & 0.001 & 0.001 & L \\ 
 62.28483 & 0.001 & 0.001 & L \\ 
 62.3 & 0.0005 & 0.0005 & L \\ 
 62.499454 & 0.001 & 0.001 & L \\ 
 62.5 & 0.001 & 0.001 & H \\ 
 62.8 & 0.0005 & 0.0005 & L \\ 
 62.8254 & 0.003 & 0.003 & H \\ 
 62.8298 & 0.003 & 0.003 & H \\ 
 63.0 & 0.001 & 0.001 & H \\ 
 63.499441 & 0.001 & 0.001 & L \\ 
 63.5 & 0.001 & 0.001 & H \\ 
 64.0 & 0.0005 & 0.0005 & H \\ 
 64.0 & 0.001 & 0.001 & H \\ 
 64.297778125 & 0.001 & 0.001 & H \\ 
 64.30116875 & 0.001 & 0.001 & H \\ 
 64.499428 & 0.001 & 0.001 & L \\ 
 64.5 & 0.001 & 0.001 & H \\ 
 65.0 & 0.001 & 0.001 & H \\ 
 65.2 & 0.0005 & 0.0005 & L \\ 
 65.499415 & 0.001 & 0.001 & L \\ 
 65.5 & 0.001 & 0.001 & H \\ 
 65.7 & 0.0005 & 0.0005 & L \\ 
 66.0 & 0.001 & 0.001 & H \\ 
 66.3719 & 0.001 & 0.001 & H \\ 
 66.3754 & 0.001 & 0.001 & H \\ 
 66.499402 & 0.001 & 0.001 & L \\ 
 66.5 & 0.001 & 0.001 & H \\ 
 66.665 & 0.001 & 0.001 & L \\ 
 67.0 & 0.001 & 0.001 & H \\ 
 67.499389 & 0.001 & 0.001 & L \\ 
 67.5 & 0.001 & 0.001 & H \\ 
 67.6 & 0.0005 & 0.0005 & L \\ 
 68.0 & 0.001 & 0.001 & H \\ 
 68.1 & 0.0005 & 0.0005 & L \\ 
 68.499376 & 0.001 & 0.001 & L \\ 
 68.5 & 0.001 & 0.001 & H \\ 
 68.6 & 0.0005 & 0.0005 & L \\ 
 69.0 & 0.001 & 0.001 & H \\ 
 69.499363 & 0.001 & 0.001 & L \\ 
 69.5 & 0.001 & 0.001 & H \\ 
 70.0 & 0.001 & 0.001 & H \\ 
 70.49935 & 0.001 & 0.001 & L \\ 
 70.5 & 0.001 & 0.001 & H \\ 
 71.0 & 0.001 & 0.001 & H \\ 
 71.0 & 0.0005 & 0.0005 & L \\ 
 71.499337 & 0.001 & 0.001 & L \\ 
 71.5 & 0.001 & 0.001 & H \\ 
 71.5 & 0.0005 & 0.0005 & L \\ 
 71.91237 & 0.0048 & 0.0024 & L \\ 
 72.0 & 0.0005 & 0.0005 & H \\ 
 72.0 & 0.001 & 0.001 & H \\ 
 72.499324 & 0.001 & 0.001 & L \\ 
 72.5 & 0.001 & 0.001 & H \\ 
 73.0 & 0.001 & 0.001 & H \\ 
 73.499311 & 0.001 & 0.001 & L \\ 
 73.5 & 0.001 & 0.001 & H \\ 
 73.9 & 0.0005 & 0.0005 & L \\ 
 74.0 & 0.001 & 0.001 & H \\ 
 74.4 & 0.0005 & 0.0005 & L \\ 
 74.5 & 0.001 & 0.001 & H \\ 
 75.0 & 0.001 & 0.001 & H \\ 
 75.5 & 0.001 & 0.001 & H \\ 
 76.0 & 0.001 & 0.001 & H \\ 
 76.3 & 0.0005 & 0.0005 & L \\ 
 76.3235 & 0.001 & 0.001 & H \\ 
 76.3235 & 0.001 & 0.001 & H \\ 
 76.411925 & 0.001 & 0.001 & H \\ 
 76.5 & 0.001 & 0.001 & H \\ 
 76.50035 & 0.001 & 0.001 & H \\ 
 76.588775 & 0.001 & 0.001 & H \\ 
 76.6772 & 0.001 & 0.001 & H \\ 
 76.75 & 0.001 & 0.001 & L \\ 
 76.765625 & 0.001 & 0.001 & H \\ 
 76.8 & 0.0005 & 0.0005 & L \\ 
 76.85405 & 0.001 & 0.001 & H \\ 
 76.942475 & 0.001 & 0.001 & H \\ 
 77.0 & 0.001 & 0.001 & H \\ 
 77.0309 & 0.001 & 0.001 & H \\ 
 77.119325 & 0.001 & 0.001 & H \\ 
 77.20775 & 0.001 & 0.001 & H \\ 
 77.296175 & 0.001 & 0.001 & H \\ 
 77.3 & 0.0005 & 0.0005 & L \\ 
 77.3846 & 0.001 & 0.001 & H \\ 
 77.473025 & 0.001 & 0.001 & H \\ 
 77.5 & 0.001 & 0.001 & H \\ 
 77.56145 & 0.001 & 0.001 & H \\ 
 77.749975 & 0.001 & 0.001 & L \\ 
 78.0 & 0.001 & 0.001 & H \\ 
 78.5 & 0.001 & 0.001 & H \\ 
 78.74995 & 0.001 & 0.001 & L \\ 
 79.0 & 0.001 & 0.001 & H \\ 
 79.2 & 0.0005 & 0.0005 & L \\ 
 79.5 & 0.001 & 0.001 & H \\ 
 79.7 & 0.0005 & 0.0005 & L \\ 
 79.749925 & 0.001 & 0.001 & L \\ 
 80.0 & 0.0005 & 0.0005 & H \\ 
 80.0 & 0.001 & 0.001 & H \\ 
 80.5 & 0.001 & 0.001 & H \\ 
 80.7499 & 0.001 & 0.001 & L \\ 
 81.0 & 0.001 & 0.001 & H \\ 
 81.5 & 0.001 & 0.001 & H \\ 
 81.749875 & 0.001 & 0.001 & L \\ 
 82.0 & 0.001 & 0.001 & H \\ 
 82.1 & 0.0005 & 0.0005 & L \\ 
 82.5 & 0.001 & 0.001 & H \\ 
 82.6 & 0.0005 & 0.0005 & L \\ 
 82.74985 & 0.001 & 0.001 & L \\ 
 83.0 & 0.001 & 0.001 & H \\ 
 83.5 & 0.001 & 0.001 & H \\ 
 83.749825 & 0.001 & 0.001 & L \\ 
 83.897765 & 0.0056 & 0.0028 & L \\ 
 84.0 & 0.001 & 0.001 & H \\ 
 84.5 & 0.001 & 0.001 & H \\ 
 84.7498 & 0.001 & 0.001 & L \\ 
 85.0 & 0.001 & 0.001 & H \\ 
 85.0 & 0.0005 & 0.0005 & L \\ 
 85.5 & 0.001 & 0.001 & H \\ 
 85.5 & 0.0005 & 0.0005 & L \\ 
 85.749775 & 0.001 & 0.001 & L \\ 
 86.0 & 0.001 & 0.001 & H \\ 
 86.5 & 0.001 & 0.001 & H \\ 
 86.74975 & 0.001 & 0.001 & L \\ 
 87.0 & 0.001 & 0.001 & H \\ 
 87.5 & 0.001 & 0.001 & H \\ 
 87.749725 & 0.001 & 0.001 & L \\ 
 87.9 & 0.0005 & 0.0005 & L \\ 
 88.0 & 0.0005 & 0.0005 & H \\ 
 88.0 & 0.001 & 0.001 & H \\ 
 88.4 & 0.0005 & 0.0005 & L \\ 
 88.5 & 0.001 & 0.001 & H \\ 
 88.7497 & 0.001 & 0.001 & L \\ 
 89.0 & 0.001 & 0.001 & H \\ 
 89.5 & 0.001 & 0.001 & H \\ 
 89.749675 & 0.001 & 0.001 & L \\ 
 90.0 & 0.001 & 0.001 & H \\ 
 90.3 & 0.0005 & 0.0005 & L \\ 
 90.5 & 0.001 & 0.001 & H \\ 
 90.74965 & 0.001 & 0.001 & L \\ 
 90.8 & 0.0005 & 0.0005 & L \\ 
 91.0 & 0.001 & 0.001 & H \\ 
 91.3 & 0.0005 & 0.0005 & L \\ 
 91.5 & 0.001 & 0.001 & H \\ 
 91.749625 & 0.001 & 0.001 & L \\ 
 92.0 & 0.001 & 0.001 & H \\ 
 92.5 & 0.001 & 0.001 & H \\ 
 92.7496 & 0.001 & 0.001 & L \\ 
 93.0 & 0.001 & 0.001 & H \\ 
 93.5 & 0.001 & 0.001 & H \\ 
 93.7 & 0.0005 & 0.0005 & L \\ 
 93.749575 & 0.001 & 0.001 & L \\ 
 94.0 & 0.001 & 0.001 & H \\ 
 94.2 & 0.0005 & 0.0005 & L \\ 
 94.2381 & 0.003 & 0.003 & H \\ 
 94.2447 & 0.003 & 0.003 & H \\ 
 94.5 & 0.001 & 0.001 & H \\ 
 94.74955 & 0.001 & 0.001 & L \\ 
 95.0 & 0.001 & 0.001 & H \\ 
 95.5 & 0.001 & 0.001 & H \\ 
 95.749525 & 0.001 & 0.001 & L \\ 
 95.88316 & 0.0064 & 0.0032 & L \\ 
 96.0 & 0.0005 & 0.0005 & H \\ 
 96.0 & 0.001 & 0.001 & H \\ 
 96.5 & 0.001 & 0.001 & H \\ 
 96.6 & 0.0005 & 0.0005 & L \\ 
 96.7495 & 0.001 & 0.001 & L \\ 
 97.0 & 0.001 & 0.001 & H \\ 
 97.1 & 0.0005 & 0.0005 & L \\ 
 97.5 & 0.001 & 0.001 & H \\ 
 97.749475 & 0.001 & 0.001 & L \\ 
 98.0 & 0.001 & 0.001 & H \\ 
 98.5 & 0.001 & 0.001 & H \\ 
 98.74945 & 0.001 & 0.001 & L \\ 
 99.0 & 0.001 & 0.001 & H \\ 
 99.0 & 0.0005 & 0.0005 & L \\ 
 99.5 & 0.001 & 0.001 & H \\ 
 99.5 & 0.0005 & 0.0005 & L \\ 
 99.749425 & 0.001 & 0.001 & L \\ 
 99.9989 & 0.001 & 0.001 & H \\ 
 100.0 & 0.001 & 0.001 & H \\ 
\hline 
\hline
\caption{Instrumental lines identified and ``cleaned'' before the \EatHs runs. The different columns represent: (I) the central frequency of the instrumental line; (II) Low-Frequency-Side (LFS) of the knockout band; (III) High-Frequency-Side (HFS) of the knockout band; (IV) the interferometer in which the instrumental lines were identified.}
\end{longtable}

\clearpage
\subsection{50-mHz signal-frequency bands where the upper limit value does not hold}
\label{A:excluded50mHzBands}
\begin{longtable}{|l|c|l|c|l|c|l|c|}
\hline
\hline
start & band &start& band& start& band& start& band\\ 
band & type & band & type & band & type & band & type \\ 
\hline
\hline
\endhead
20.40 & M  & 20.90 & D  & 20.80 & M  & 21.45 & M \\ 
22.40 & D  & 23.90 & M  & 24.45 & D  & 24.20 & M \\ 
25.25 & M  & 25.60 & M  & 26.05 & M  & 26.90 & M \\ 
27.50 & D  & 27.45 & M  & 27.85 & D  & 27.55 & M \\ 
28.55 & D  & 28.90 & M  & 29.15 & D  & 30.60 & D \\ 
30.85 & M  & 31.10 & M  & 31.40 & I  & 31.75 & M \\ 
32.35 & M  & 32.90 & M  & 33.05 & M  & 34.80 & M \\ 
34.60 & C  & 34.65 & C  & 34.70 & C  & 34.75 & C \\ 
35.20 & C  & 35.25 & C  & 35.30 & C  & 35.35 & C \\ 
35.70 & M  & 35.80 & C  & 35.85 & C  & 35.90 & C \\ 
35.95 & C  & 36.60 & M  & 36.60 & C  & 36.65 & C \\ 
36.70 & C  & 36.75 & C  & 37.25 & M  & 39.75 & M \\ 
40.20 & M  & 40.85 & D  & 42.80 & M  & 43.65 & M \\ 
44.70 & D  & 44.65 & M  & 45.30 & D  & 45.35 & M \\ 
46.90 & M  & 47.65 & M  & 48.95 & M  & 50.25 & M \\ 
51.00 & M  & 52.30 & M  & 52.60 & M  & 52.80 & I \\ 
53.05 & M  & 54.70 & M  & 55.05 & M  & 55.60 & M \\ 
56.80 & D  & 57.10 & M  & 58.95 & M  & 59.50 & M \\ 
59.55 & M  & 59.95 & C  & 60.00 & C  & 61.00 & M \\ 
61.05 & M  & 62.45 & D  & 62.05 & M  & 66.65 & M \\ 
74.50 & D  & 74.45 & M  & 75.00 & M  & 76.60 & D \\ 
76.65 & M  & 83.30 & M  & 85.80 & M  & 89.40 & D \\ 
89.35 & M  & 90.00 & M  & 99.95 & D  & & \\
\hline
\hline
\caption{50-mHz search-frequency bands that are excluded from the results. Bands are excluded from the results if they were identified as disturbed
based on visual inspection (D), if they were identified as mildly disturbed based on visual inspection then excluded later in the analysys (M), if they contained a hardware injection (I) or where the results were produced from entirely fake data as detailed in Table I (C). Bands labelled D, C or I are excluded from the analysis. }
\end{longtable}

\clearpage

\iftoggle{endauthorlist}{
  %
  %
  \let\author\myauthor
  \let\affiliation\myaffiliation
  \let\maketitle\mymaketitle
  \title{Authors}
  \pacs{}
  
  \newpage
  \maketitle
}

\end{document}